\documentclass[11pt]{article} 
\usepackage{amsmath}
\usepackage{amsfonts,amssymb}

\begin{document}

\newcommand{\nl}{\nonumber\\}
\newcommand{\nnl}{\nl[6mm]}
\newcommand{\nle}{\nl[-2.5mm]\\[-2.5mm]}
\newcommand{\nlb}[1]{\nl[-2.0mm]\label{#1}\\[-2.0mm]}
\newcommand{\ab}{\allowbreak}

\renewcommand{\leq}{\leqslant}              
\renewcommand{\geq}{\geqslant}

\renewcommand{\theequation}{\thesection.\arabic{equation}}
\let\ssection=\section
\renewcommand{\section}{\setcounter{equation}{0}\ssection}

\newcommand{\be}{\bes}
\newcommand{\ee}{\ees}
\newcommand{\bes}{\begin{eqnarray}}
\newcommand{\ees}{\end{eqnarray}}
\newcommand{\eens}{\nonumber\end{eqnarray}}
\newcommand{\barr}{\begin{array}}
\newcommand{\earr}{\end{array}}

\renewcommand{\/}{\over}
\renewcommand{\d}{\partial}
\newcommand{\Dslash}{\hbox{$D\kern-2.4mm/\,$}}
\newcommand{\dd}[1]{\ab\delta/\delta {#1}}
\newcommand{\ddt}{{d\/dt}}

\newcommand{\no}[1]{{\,:\kern-0.7mm #1\kern-1.2mm:\,}}

\newcommand{\mm}{{\mathbf m}}
\newcommand{\nn}{{\mathbf n}}
\newcommand{\cmm}{{,\mm}}
\newcommand{\cnn}{{,\nn}}

\newcommand{\qsmu}{q^*_\mu}
\newcommand{\qsnu}{q^*_\nu}
\newcommand{\psmu}{p_*^\mu}

\newcommand{\half}{{1\/2}}
\newcommand{\quart}{{1\/4}}
\newcommand{\tpi}{{1\/2\pi i}}
\newcommand{\bra}[1]{\big{\langle}#1\big{|}}
\newcommand{\ket}[1]{\big{|}#1\big{\rangle}}

\newcommand{\da}{\d_\alpha}
\newcommand{\db}{\d_\beta}
\newcommand{\dc}{\d_\gamma}
\newcommand{\za}{\zeta_a}
\newcommand{\zb}{\zeta_b}
\newcommand{\zc}{\zeta_c}
\newcommand{\Mab}{M^{\alpha\beta}}
\newcommand{\Kab}{K_{\alpha\beta}}

\newcommand{\fa}{\phi^\alpha}
\newcommand{\fb}{\phi^\beta}
\newcommand{\pa}{\pi_\alpha}
\newcommand{\pb}{\pi_\beta}
\newcommand{\Ea}{\EE_\alpha}
\newcommand{\Eb}{\EE_\beta}

\newcommand{\fs}{\phi^*}
\newcommand{\ps}{\pi_*}
\newcommand{\fsa}{\phi^*_\alpha}
\newcommand{\fsb}{\phi^*_\beta}
\newcommand{\fsc}{\phi^*_\gamma}
\newcommand{\psa}{\pi_*^\alpha}
\newcommand{\psb}{\pi_*^\beta}
\newcommand{\fsi}{\phi^*_i}

\newcommand{\fA}{\phi^A}
\newcommand{\fB}{\phi^B}
\newcommand{\pA}{\pi_A}
\newcommand{\wfA}{\bar\phi^A}
\newcommand{\wfB}{\bar\phi^B}

\newcommand{\As}{A_*}
\newcommand{\Es}{E^*}
\newcommand{\wA}{{\bar A}}
\newcommand{\wE}{{\bar E}}
\newcommand{\wAs}{{\bar A}{}_*}
\newcommand{\wEs}{{\bar E}{}^*}
\newcommand{\wc}{{\bar c}}
\newcommand{\wzeta}{{\bar\zeta}}
\newcommand{\wF}{{\bar F}}

\newcommand{\wf}{{\bar \phi}}
\renewcommand{\wp}{{\bar \pi}}
\newcommand{\wfs}{{\bar \phi}{}^*}
\newcommand{\wps}{{\bar \pi}_*}

\newcommand{\fm}{\phi_\cmm}
\newcommand{\fn}{\phi_\cnn}
\newcommand{\pim}{\pi^\cmm}
\newcommand{\pin}{\pi^\cnn}
\newcommand{\fsm}{\phi^*_{c\mm}}
\newcommand{\fsn}{\phi^*_\cnn}
\newcommand{\psm}{\pi_*^\cmm}
\newcommand{\psn}{\pi_*^\cnn}

\renewcommand{\fam}{\phi^\alpha_\cmm}
\newcommand{\fbn}{\phi^\beta_\cnn}
\newcommand{\pam}{\pi_\alpha^\cmm}
\newcommand{\pbn}{\pi_\beta^\cnn}
\newcommand{\zam}{\zeta_{a\cmm}}
\newcommand{\zcm}{\zeta_{c\cmm}}
\newcommand{\cam}{c^a_\cmm}
\newcommand{\wfam}{{\bar \phi}{}^\alpha_\cmm}
\newcommand{\wfbn}{{\bar \phi}{}^\beta_\cnn}
\newcommand{\wpam}{{\bar \pi}{}_\alpha^\cmm}

\newcommand{\fAm}{\fA_\cmm}
\newcommand{\fAn}{\fA_\cnn}
\newcommand{\fBn}{\fB_\cnn}
\newcommand{\pAm}{\pA^\cmm}
\newcommand{\pAn}{\pA^\cnn}
\newcommand{\wfAm}{\bar\fA_\cmm}

\newcommand{\wfm}{{\bar \phi}_\cmm}
\newcommand{\wfn}{{\bar \phi}_\cnn}
\newcommand{\wpm}{{\bar \pi}^\cmm}
\newcommand{\wpsm}{{\bar \ps}^\cmm}

\newcommand{\wfa}{{\bar \phi}{}^\alpha}

\newcommand{\fsam}{\phi^*_{\alpha,\mm}}
\newcommand{\fsbn}{\phi^*_{\beta,\nn}}
\newcommand{\psam}{\pi_*^{\alpha,\mm}}
\newcommand{\psbn}{\pi_*^{\beta,\nn}}
\newcommand{\Eam}{\EE_{\alpha,\mm}}
\newcommand{\wfsam}{\bar\phi^*_{\alpha,\mm}}

\newcommand{\dam}{\d_\alpha^\mm}
\newcommand{\dbn}{\d_\beta^\nn}

\newcommand{\si}{\sigma}
\newcommand{\eps}{\epsilon}
\newcommand{\dlt}{\delta}
\newcommand{\om}{\omega}
\newcommand{\al}{\alpha}
\newcommand{\bt}{\beta}
\newcommand{\gm}{\gamma}
\newcommand{\ka}{\kappa}
\newcommand{\la}{\lambda}
\newcommand{\vth}{\vartheta}
\renewcommand{\th}{\theta}
\newcommand{\rep}{\varrho}

\newcommand{\vect}{{\mathfrak{vect}}}
\newcommand{\map}{{\mathfrak{map}}}

\newcommand{\im}{{\rm im}\ }
\newcommand{\ext}{{\rm ext}\ }
\newcommand{\e}{{\rm e}}
\renewcommand{\div}{{\rm div}}
\newcommand{\afn}{{\rm afn\,}}
\newcommand{\til}{{\tilde{\ }}}

\newcommand{\eikx}{\e^{ik\cdot x}}
\newcommand{\dNx}{{d^N \kern-0.2mm x\ }}
\newcommand{\dNxp}{{d^N \kern-0.2mm x'\ }}
\newcommand{\dNxb}{{d^N \kern-0.2mm x''\ }}
\newcommand{\dNk}{{d^N \kern-0.2mm k\ }}
\newcommand{\dNkp}{{d^N \kern-0.2mm k'\ }}

\newcommand{\summ}[1]{\sum_{|\mm|\leq #1}}
\newcommand{\sumn}[1]{\sum_{|\nn|\leq #1}}
\newcommand{\summnp}{\sum_{|\mm|\leq|\nn|\leq p }}
\newcommand{\sumnmp}[1]{\sum_{|\nn|\leq|\mm|\leq p#1 }}
\newcommand{\summp}{\summ{p}}

\newcommand{\dmu}{{\d_\mu}}
\newcommand{\dnu}{{\d_\nu}}

\newcommand{\larroww}[1]{{\ \stackrel{#1}{\longleftarrow}\ }}
\newcommand{\rarroww}[1]{{\ \stackrel{#1}{\longrightarrow}\ }}
\newcommand{\intdm}{\int_{-\infty}^\infty dm\ }

\newcommand{\mndmn}{{\mm\choose\nn}\d_{\mm-\nn}}

\newcommand{\tr}{{\rm tr}}
\newcommand{\oj}{{\mathfrak g}}
\newcommand{\hh}{{\mathfrak h}}
\newcommand{\uu}{{\mathfrak u}}

\newcommand{\J}{{\cal J}}
\newcommand{\D}{{\cal D}}
\newcommand{\U}{{\cal U}}
\newcommand{\N}{{\cal N}}
\newcommand{\OO}{{\cal O}}
\newcommand{\QQ}{{\cal Q}}
\newcommand{\PP}{{\cal P}}
\newcommand{\EE}{{\cal E}}
\newcommand{\FF}{{\cal F}}
\newcommand{\HH}{{\cal H}}
\newcommand{\II}{{\cal I}}
\newcommand{\GG}{{\cal G}}

\newcommand{\cl}{{cl}}
\newcommand{\qm}{{qm}}
                                            
\newcommand{\TT}{{\mathbb T}}
\newcommand{\RR}{{\mathbb R}}
\newcommand{\CC}{{\mathbb C}}
\newcommand{\ZZ}{{\mathbb Z}}
\newcommand{\NN}{{\mathbb N}}

\title{{Manifestly covariant canonical quantization II: Gauge theory
and anomalies}}

\author{T. A. Larsson \\
Vanadisv\"agen 29, S-113 23 Stockholm, Sweden\\
email: thomas.larsson@hdd.se}

\maketitle 
\begin{abstract} 
In {\tt hep-th/0411028} a new manifestly covariant canonical
quantization method was developed. The idea is to quantize in the 
phase space of arbitrary histories first, and impose dyna\-mics as
first-class constraints afterwards. The Hamiltonian is defined
covariantly as the generator of rigid translations of the fields
relative to the observer. This formalism is now applied to theories
with gauge symmetries, in particular electromagnetism and Yang-Mills
theory. The gauge algebra acquires an abelian extension proportional
to the quadratic Casimir operator. Unlike conventional gauge anomalies
proportional to the third Casimir, this is not inconsistent. On the
contrary, a well-defined and non-zero charge operator is only
compatible with unitarity in the presence of such anomalies. This
anomaly is invisible in field theory because it is a functional of the
observer's trajectory, which is conventionally ignored.
\end{abstract}

\vskip 3 cm
PACS (2003): 02.20.Tw, 03.65.Ca, 03.70.+k, 11.10.Ef.


\newpage

{\em
Dedicated to the memory of Julia Teng\aa\ and her parents,
who disappeared in the Wave at Khao Lak. }

\vskip 1cm

\section{Introduction} 

In the first paper in this series \cite{Lar04}, we introduced a new
canonical quantization method, which preserves manifest covariance. It
is based on two key ideas: regard the Euler-Lagrange (EL) equations as
first-class constraints in the phase space of arbitary histories, and
expand all fields in a Taylor series around the observer's trajectory in
spacetime. The real motivation for introducing a new way to look at the
old quantum theory becomes apparent in the present paper, where the
formalism is applied to systems with constraints and gauge symmetries.
{\em We demand that the constraint algebra be realized as well-defined
operators on the kinematical Hilbert space.} A first step in this
direction is to understand the quantum representation theory of the
constraint algebra. Fortunately, in the last decade much has been learnt
about the representations of algebras of diffeomorphism and gauge 
transformations in more than one dimension 
\cite{BBS01,Lar98,Lar02,Lar03,RM94}.
The time is now ripe to apply these insights to physics.

It is known but not always appreciated that the notion of gauge
invariance is problematic on the quantum level already in QED. E.g., a
recent article \cite{Stei04} begins: ``This article is concerned with a
major unsolved problem of QED, that of an exact formulation of gauge
invariance, more especially the problem of an exact characterization of
gauge transformations and their uses.''
The problem with gauge transformations on the quantum level is
due to a real obstruction: in all unitary lowest-energy
representations, the groups of gauge transformations and
diffeomorphisms acquire quantum corrections. This is well known in one
dimension, where the relevant algebras are the affine Kac-Moody and
Virasoro algebras, respectively, but the same thing is true in higher
dimensions as well.

Let us briefly outline the main ideas in this paper.
Phase space is a covariant concept; it is the space of histories which
solve the dyna\-mics. Each phase space point $(q,p)$ generates a unique
history $(q(t),p(t))$ under Hamiltonian evolution. We may choose to
coordinatize phase space by $(q,p) = (q(0),p(0))$, but this is only a
choice of coordinates, and physics is of course independent of this choice.
We make the space of arbitrary histories 
$(q(t),p(t))$ into a phase space $\PP$ by defining the Poisson brackets
\be
[p(t), q(t')] = \dlt(t-t'), \qquad 
[p(t), p(t')] = [q(t), q(t')] = 0.
\label{histbrack}
\ee
The EL equations now define a constraint $\EE(t)\approx0$ in $\PP$;
since $\EE(t)$ only depends on $q(t)$ this constraint is first class.
This observation allows us to apply powerful cohomological methods from
BRST quantization of theories with first class constraints. In other
words, the idea is to quantize in the history phase space first and
to impose dyna\-mics afterwards, by passing to cohomology. Since dyna\-mics
is regarded as a constraint, this is non-trivial even for systems without
gauge symmetries, like the harmonic oscillator and the free scalar field
treated in \cite{Lar04}.

The second idea is to expand all fields in a Taylor series around the
observer's trajectory $q^\mu(t)$. Since covariance only makes sense
for field theories, we now denote the histories by 
$(\phi(x), \pi(x))$, and define 
\be
\phi(x) = \sum_\mm {1\/\mm!} \phi_\cmm(t)(x-q(t))^\mm. 
\ee 
One reason for describing the history phase space with Taylor
coordinates $(\phi_\cmm(t),\ab q^\mu(t),\ab \pi^\cmm(t),\ab
p_\mu(t))$ rather than field coordinates $(\phi(x),\pi(x))$ 
is that we can write down a
covariant expression for the Hamiltonian; the operator 
\be
H = i\int dt \no{ \dot q^\mu(t)p_\mu(t) } 
\ee 
translates the observer relative
to the fields or vice versa. In particular, if $q^\mu(t) = u^\mu t$
where $u^\mu = (1,0,0,0)$, then the Hamiltonian acts on the fields as
expected:
\be 
[H, \phi(x)] = -i{\d\/\d x^0}\phi(x). 
\ee 
Taken together, these two ideas yield a formulation of
quantum theory which is both canonical and covariant. This is
aesthetetically pleasing, but this is not really a good enough reason to
motivate a new formalism. After all, decades of work have gone into path
integrals and conventional canonical quantization, and those methods
yield accurate predictions for observed quantities. In contrast, the only
things that have been calculated so far with the present method are the
spectra of the harmonic oscillator and the free scalar field
\cite{Lar04}. The true motivation is the possibility to realize the 
gauge generators as well-defined operators.

Let us define what it means for an operator to be well-defined.
Ideally, we want unitary and perhaps bounded operators acting on
a Hilbert space. However, nothing is said about unitarity or the Hilbert
space norm in this paper. Instead, we use the weaker notion that
an operator acting on a linear space
with countable basis is well-defined if it takes a basis state into a
{\em finite} linear combination of basis states. In particular, the
brackets of two operators $[A,B]$ must itself be an operator, and an
infinite constant times the unit operator is not well-defined.

In canonical quantization of low-dimensional systems, the constraint
algebra is realized as well-defined operators on the kinematical Hilbert
space. This is satisfied e.g. in first-quantized string theory, where 
the constraint
algebra of Weyl rescalings is realized as the Virasoro operators
\be
L_m = \sum_n \no{a_n a_{m-n}},
\ee
which act on the kinematical Hilbert space
generated by $a_m$'s with $m\geq0$. However, in conventional canonical
quantization of higher-dimensional
theories such as Yang-Mills theory and general relativity in four
dimensions, the quantum constraints are not well-defined operators,
because infinities arise which can not be removed by normal ordering.

The relevant constraint algebras can typically be identified as
algebras of gauge transformations or diffeomorphisms in four
dimensions, whose natural habitat is a linear space built from a
truncation of the history variables. More precisely, the space of
Taylor coefficients with multi-indices of length $|\mm|\leq p$
carries a nonlinear realization of these algebras. After
quantization, we obtain a lowest-energy representation on a linear
space. The representation is projective and has an appropriate
concept of energy which is bounded from below. By a projective
representation we mean that the constraint algebra acquires an
extension, i.e. it becomes a higher-dimensional generalization of the
affine Kac-Moody or the Virasoro algebras. It is hardly surprising
that out formalism yields well-defined representations of such
extended algebras, since it grew out of the representation theory of
toroidal Lie algebras.

An extension of a gauge algebra can be considered as a gauge anomaly,
which usually is considered to be inconsistent \cite{NAG85}. However,
this is a far too simplistic point of view. A gauge symmetry is a
redundancy of the description, and an anomaly therefore means that
some gauge degrees of freedom become physical at the quantum level
(quantum-mechanical gauge symmetry breaking).
The anomalous theory {\em may} be inconsistent, but does not have to
be so provided that the anomalous representation of the gauge algebra
is unitary. If there is an anomaly, the gauge symmetry is of course
no longer gauge, but must be treated as an anomalous global symmetry.
Moreover, without a gauge anomaly it is really impossible to have
non-zero electric charge, because the charge operator is a gauge
generator. Conventionally, this problem is ignored because gauge
generators are not well-defined operators.

Since this view of gauge
anomalies is unorthodox, Section \ref{sec:anom} is devoted to a longer
discussion of this topic, together with a general description of the BRST
approach to quantization. That section also contains an explicit 
description of the Lie algebra extensions on the $N$-dimensional torus,
making the connection to affine and Virasoro algebras very clear.
It should be emphasized that the anomalies encountered here are of a
new type, not visible in field theory, because the relevant cocycles are
functionals of the observer's trajectory. 
Another important property is that they do not ruin the Noether
identities; charge conservation is implemented as an operator
equation, as explained in Section \ref{sec:fermions}.

The philosophy in this paper is in many ways similar to the History
Projection Operator formalism developed by Savvidou and others.
In particular, the Poisson brackets (\ref{histbrack}) in the history
phase space were first introduced in \cite{Sav98}. It was emphasized in
\cite{Sav04} that conventional canonical quantization leads to severe
conceptual difficulties when applied to covariant theories, especially
gravity. A substantial difference is that the observer's trajectory
was not explicitly introduced in those papers, and hence no 
observer-dependent anomalies were found.

The rest of the paper is organized as follows. Section \ref{sec:KT}
contains a general description of the cohomological approach to
quantization. We introduce antifields and ghosts and
obtain the reduced phase space in the BRST cohomology. However, the
gauge generators and the BRST operator become ill defined after 
quantization. To remedy that, we regularize the theory in 
Section \ref{sec:covar} by passing to $p$-jet space. The unique feature
with this regularization is that the full gauge symmetry is manifest in
the truncated theory. After quantization, the gauge generators remain
well defined but the brackets may be anomalous, i.e. the BRST operator
typically fails to be nilpotent. The BRST operator is a sum of two
terms, $Q_{BRST} = Q_{KT} + Q_{Long}$, and the Koszul-Tate (KT) part
$Q_{KT}$ is still nilpotent. This means that we must use the KT
cohomology rather
than the BRST cohomology to construct the physical Hilbert space.
In Section \ref{sec:Gauge} we describe the quantum (normal-ordered)
form of the gauge generators and their associated anomalies.

In Section \ref{sec:Maxwell} the formalism is applied to the free
Maxwell field. This case is special because it is anomaly free; the
Kac-Moody-like cocycle is proportional to the second Casimir, which
vanishes in the adjoint representation of $\uu(1)$. Standard results
are recovered, in particular photons with two physical polarizations.
There is some spurious cohomology because the Hessian is singular,
but this phenomenon occurred already for the harmonic oscillator
\cite{Lar04} and has nothing to do with gauge invariance.

In the next section we couple a charged Dirac field to the Maxwell
field. The main novelty here is the presence of a non-zero gauge
anomaly. The KT operator still ensures that only the gauge-invariant
field strength $F_{\mu\nu}$ survives in cohomology, but now there are
three photon polarizations; in the interacting theory the photons are
virtual, and virtual photons may have unphysical polarizations.

In Section \ref{sec:Yang-Mills} a pure non-abelian Yang-Mills field is
considered. Here the second Casimir is non-zero already without fermions,
because the gluons themselves carry color charge, and accordingly the
Yang-Mills algebra acquires quantum corrections. The final section
contains a brief conclusion.

\section{ Nilpotency, anomalies, and consistency }
\label{sec:anom}

The cleanest way to deal with gauge symmetries is through the BRST
formalism. For each gauge symmetry generator $J_a$, we introduce a
ghost $c^a$ and ghost momentum $b_a$, and define the BRST operator
\be
Q = J_a c^a - \half f_{ab}{}^c c^a c^b b_c.
\ee
The classical BRST operator is nilpotent, 
$Q^2 = \{Q,Q\} = 0$,
where $\{\cdot,\cdot\}$ is the symmetric bracket.
A classical observable $\OO$ is BRST closed, $[Q, \OO] = 0$,
and two observables are equivalent if they differ by a BRST exact term,
$\OO \sim \OO + [Q, F]$. In other words, the space of functions
on the extended phase space, including ghosts and ghost momenta,
decomposes into subspaces of fixed ghost number,
$C(\PP^*) = \sum_k C^k$, such that
\be
\ldots C^{-2} \rarroww{Q} C^{-1} \rarroww{Q} C^0 
\rarroww{Q} C^1 \rarroww{Q} 
C^2 \rarroww{Q} \ldots.
\label{complex0}
\ee
The space of classical observables can be identified with the zeroth
cohomology group of this complex,
\be
H^0(Q) = {(\ker Q)_0\/(\im Q)_0},
\ee
whereas the higher cohomology groups are zero; we say that the complex
(\ref{complex0}) gives a resolution of the physical (gauge-invariant)
phase space.

After quantization, we obtain two cohomologies. In the state
cohomology, a physical state is BRST closed, $Q\ket{phys} = 0$, and
two states are equivalent if they differ by an exact term,
$\ket{phys} \sim \ket{phys} + Q\ket{}$. In the operator cohomology, a
physical operator is BRST closed, $[Q, \OO] = 0$, and two operators
are equivalent if they differ by a BRST exact term, $\OO \sim \OO +
[Q, F]$. It is a standard result that these definitions are
compatible.

One thing may go wrong upon quantization: the gauge symmetry
may acquire quantum corrections, i.e. an anomaly. The gauge algebra 
then assumes the form
\be
[J_a, J_b] = f_{ab}{}^c J_c + D_{ab},
\label{ext}
\ee
and the quantum BRST operator ceases to be nilpotent,
$\{Q,Q\} \sim D_{ab}$. Then it makes no sense to
require that $Q\ket{phys} = 0$, because it would mean that
\be
\{Q,Q\}\ket{phys} \sim D_{ab}\ket{phys} = 0.
\ee
In particular, if $D_{ab}$ is an invertible operator this would mean
that $\ket{phys}=0$, i.e. that there are no physical states at all.

However, it must be realized that the extension (\ref{ext}) is not
necessarily inconsistent by itself. The inconsistency was our requirement
that all physical states be annihilated by the no longer nilpotent BRST
operator. It is quite conceivable that the unreduced Hilbert space has a
positive-definite norm which is preserved by the algebra (\ref{ext}) and
by time evolution. This happens in the chiral Schwinger model \cite{JR85}
and perhaps also in Liouville field theory. More importantly,
anomalous conformal symmetry plays an important role in two-dimensional
statistical physics \cite{FMS96}. In fact, it is well known in this
context that infinite-dimensional spacetime symmetries (gauge or not)
are compatible with locality, in the sense of correlation functions 
which depend on separation, only in the presence of an anomaly.

In the situation that we are interested in, the BRST operator can be
written as $Q \equiv Q_{BRST} = Q_{KT} + Q_{Long}$, where the Koszul-Tate (KT)
operator $Q_{KT}$ always remains nilpotent after quantization, but the
longitudinal part $Q_{Long}$ may become anomalous. Then any state of
the form $Q_{KT}\ket{}$ can and must be modded out. This is
done by identifying the physical Hilbert space with the zeroth KT
cohomology group. There are thus two different situations:
\begin{enumerate}
\item
$Q^2_{BRST} = 0$. Then $\HH_{phys} = H^0(Q_{BRST})$, and the gauge
degrees of freedom are truly redundant, even after quantization.
\item
$Q^2_{BRST} \neq 0$ but $Q^2_{KT} = 0$. Then 
$\HH_{phys} = H^0(Q_{KT})$, and the gauge degrees of freedom become
physical after quantization.
\end{enumerate}
In either case one must check that the final theory is consistent, i.e.
that $\HH_{phys}$ has a positive-definite norm which is preserved by
time evolution and all symmetries. This condition may or may not fail.

It is common to instinctively reject all gauge anomalies as
inconsistent, due to experience with the standard model. It is
therefore important to point out that the gauge anomalies considered
in the present paper are of a different type than conventional ones.
In field theory, anomalies typically arise when chiral fermions are
coupled to gauge fields, and they are proportional to the third
Casimir operator \cite{Bon86,NAG85}. In contrast, the gauge anomalies
described here are functionals of the observer's trajectory in
spacetime, and they are proportional to the second Casimir. In
particular, they only vanish for the pure Maxwell field. Moreover,
within field theory no gravitational anomalies at all exist in four
dimensions, but a generalization of the Virasoro algebra certainly
exists in any number of dimensions, and this extension arises upon
quantization.

To make contact with the Virasoro algebra in its most familiar form, we
describe its multi-dimensional sibling in a Fourier basis on the
$N$-dimensional torus. Recall first that the algebra of diffeomorphisms on
the circle, $\vect(1)$, has generators
\be
L_m = -i \exp(imx) {d\/d x},
\label{Lm}
\ee
where $x \in S^1$.
$\vect(1)$ has a central extension, known as the Virasoro algebra:
\be
[L_m, L_n] = (n-m)L_{m+n} - {c\/12} (m^3-m) \delta_{m+n},
\label{Vir}
\ee
where $c$ is a c-number known as the central charge or conformal
anomaly. This means that the Virasoro algebra is a Lie 
algebra; anti-symmetry and the Jacobi identities still hold. The
term linear in $m$ is unimportant, because it can be removed by a
redefinition of $L_0$. The cubic term $m^3$ is a non-trivial 
extension which cannot be removed by any redefinition.

The generators (\ref{Lm}) immediately generalize to vector fields on the
$N$-dimensional torus:
\be
L_\mu(m) = -i \exp(i m_\rho x^\rho) \dmu,
\ee
where $x = (x^\mu)$, $\mu = 1, 2, ..., N$ is a point in 
$N$-dimensional space and $m = (m_\mu)$. The Einstein convention is used;
repeated indices, one up and one down, are implicitly summed over. 
These operators generate the algebra $\vect(N)$:
\be
[L_\mu(m), L_\nu(n)] = n_\mu L_\nu(m+n) - m_\nu L_\mu(m+n).
\ee
The question is now whether the Virasoro extension, i.e. the 
$m^3$ term in (\ref{Vir}), also generalizes to higher dimensions.

Rewrite the ordinary Virasoro algebra (\ref{Vir}) as
\bes
[L_m, L_n] &=& (n-m)L_{m+n} + c m^2 n S_{m+n}, \nl
{[}L_m, S_n] &=& (n+m)S_{m+n}, \nle
{[}S_m, S_n] &=& 0, \nl
m S_m &=& 0.
\eens
It is easy to see that the two formulations of $Vir$ are equivalent
(I have absorbed the linear cocycle into a redefinition of $L_0$).
The second formulation immediately generalizes to $N$ dimensions.
The defining relations are
\bes
[L_\mu(m), L_\nu(n)] &=& n_\mu L_\nu(m+n) - m_\nu L_\mu(m+n) \nl 
&&  + (c_1 m_\nu n_\mu + c_2 m_\mu n_\nu) m_\rho S^\rho(m+n), \nl
{[}L_\mu(m), S^\nu(n)] &=& n_\mu S^\nu(m+n)
 + \delta^\nu_\mu m_\rho S^\rho(m+n), 
\nlb{mVir}
{[}S^\mu(m), S^\nu(n)] &=& 0, \nl
m_\mu S^\mu(m) &=& 0.
\eens
This is an extension of $\vect(N)$ by the abelian ideal with basis 
$S^\mu(m)$. 
Geometrically, we can think of $L_\mu(m)$ as a vector field 
and $S^\mu(m) = \eps^{\mu\nu_2..\nu_N} \ab S_{\nu_2..\nu_N}(m)$ 
as a dual one-form (and $S_{\nu_2..\nu_N}(m)$ as an $(N-1)$-form);
the last condition expresses closedness. 
The cocycle proportional to $c_1$ was discovered by 
Rao and Moody \cite{RM94}, and the one proportional to $c_2$ by
this author \cite{Lar91}. 

There is also a similar multi-dimensional generalization of affine
Kac-Moody algebras, presumably first written down by Kassel
\cite{Kas85}. It is sometimes called the central extension, but this
term is somewhat misleading because the extension does not commute with
diffeomorphisms, although it does commute with all gauge
transformations.

Let $\oj$ be a finite-dimensional Lie algebra with structure
constants $f_{ab}{}^c$ and Killing metric $\delta_{ab}$. The Kassel
extension of the current algebra $\map(N,\oj)$ is defined by the
brackets
\bes
[J_a(m), J_b(n)] &=& f_{ab}{}^c J_c(m+n)
  + k \delta_{ab} m_\rho S^\rho(m+n), \nl
{[}J_a(m), S^\mu(n)] &=& [S^\mu(m), S^\nu(n)] = 0, 
\label{affine}\\
m_\mu S^\mu(m) &\equiv& 0.
\eens
This algebra admits an intertwining action of $Vir(N)$:
\be
[L_\mu(m), J_a(n)] = n_\mu J_a(m+n).
\ee

The current algebra $\map(N,\oj)$ also admits another type of
extension in some dimensions. The best known example is the
Mickelsson-Faddeev algebra, relevant for the conventional
anomalies in field theory, which arise when chiral fermions are
coupled to gauge fields in three spatial dimensions. Let
$d_{abc} = \tr \{T_a, T_b\}T_c$ be the totally symmetric third
Casimir operator, and let $\eps^{\mu\nu\rho}$ be the totally
anti-symmetric epsilon tensor in three dimensions. The 
Mickelsson-Faddeev algebra \cite{Mi89} reads in a Fourier basis:
\bes
[J_a(m), J_b(n)] &=& f_{ab}{}^c J_c(m+n)
  + d_{abc} \epsilon^{\mu\nu\rho} m_\mu n_\nu A^c_\rho(m+n), \nl
{[}J_a(m), A^b_\nu(n)] &=& f_{ab}{}^c A^c_\nu(m+n) 
  + \delta_a^b m_\nu \dlt(m+n), 
\label{MF}\\
{[}A^a_\mu(m), A^b_\nu(n)] &=& 0.
\eens
$A^a_\mu(m)$ are the Fourier components of the gauge connection.

Note that $Q_a \equiv J_a(0)$ generate a Lie algebra isomorphic to
$\oj$, whose Cartan subalgebra is identified with the charges.
Moreover, the subalgebra of (\ref{affine}) spanned by 
$J_a(m_0) \equiv J_a(m)$, where $m = (m_0, 0, ..., 0) \in \ZZ$, reads
\be
[J_a(m_0), J_b(n_0)] &=& f_{ab}{}^c J_c(m_0+n_0)
+ k \delta_{ab} m_0 \dlt(m_0+n_0),
\ee
which we recognize as the affine algebra $\widehat{\oj}$. Since all
non-trivial unitary irreps of $\widehat{\oj}$ has $k>0$ \cite{GO86}, 
it is impossible to combine unitary and non-zero $\oj$ charges also
for the higher-dimensional algebra (\ref{affine}).
This follows immediately from the fact that the restriction of a 
unitary irrep to a subalgebra is also unitary (albeit in general
reducible).
Hence, rather than being inconsistent, the anomaly in (\ref{affine})
is indeed necessary for consistently including charge. This problem
is circumvented in field theory simply because e.g. the electric
charge is not a well-defined operator \cite{Stei04}.

In contrast, the Mickelsson-Faddeev algebra (\ref{MF}) has apparently 
no faithful unitary representations on a separable Hilbert space
\cite{Pic89}, which presumably means that that it
should be avoided. Indeed, Nature appears to abhor this kind of
anomaly, which is proportional to the third Casimir.

\section{Koszul-Tate and BRST cohomologies for constrained systems}
\label{sec:KT}

The Hamiltonian formulation of quantum theory is more fundamental
than the Lagrangian one, but it breaks manifest covariance. This is a
major disadvantage, especially for theories with local symmetries. A
manifestly covariant Hamiltonian formulation in the absense of gauge
symmetries was described in \cite{Lar04}, inspired by the
Batalin-Vilkovisky or antifield approach \cite{HT92}. The idea is to
consider the space of arbitrary histories and their momenta as a
phase space, and dyna\-mics, i.e. the Euler-Lagrange (EL) equations,
as first class constraints. This enables us to apply BRST techniques
in the history phase space; quantize first and impose dyna\-mics
afterwards. Let us review this idea and then extend it to theories
with irreducible gauge symmetries.

Consider a dynamical system described by a set of fields $\fa$
and an action $S$. The
EL equations read
\be
\Ea =\da S \equiv {\dlt S\/\dlt \fa} = 0.
\label{EL}
\ee
We introduce an antifield $\fsa$ for each EL equation $\Ea=0$, and
replace the space of $\phi$-histories $\QQ$ by the extended history
space $\QQ^*$, spanned by both $\phi$ and $\fs$.
In $\QQ^*$ we define the Koszul-Tate (KT) differential $\dlt$ by
\bes
\dlt \fa &=& 0, 
\nlb{dlt0}
\dlt \fsa &=& \Ea.
\eens
One checks that $\dlt$ is nilpotent, $\dlt^2 = 0$.
Define the antifield number $\afn \fa = 0$, $\afn \fsa = 1$. The KT
differential clearly has antifield number $\afn\dlt = -1$.

The function space $C(\QQ^*)$ decomposes into subspaces $C^k(\QQ^*)$ of
fixed antifield number
\be
C(\QQ^*) = \sum_{k=0}^\infty C^k(\QQ^*)
\ee
The KT differential makes  $C(\QQ^*)$ into a differential complex,
\be
0 \larroww \dlt C^0 \larroww \dlt C^1 \larroww \dlt 
C^2 \larroww \dlt \ldots
\label{complex1}
\ee
The cohomology spaces are defined as usual by
$H_\cl^\bullet(\dlt) = \ker\dlt/\im\dlt$, i.e.
$H_\cl^k(\dlt) = (\ker\dlt)_k/(\im\dlt)_k$, where
the subscript $\cl$ indicates that we deal with a classical phase space.
It is easy to see that
\bes
(\ker \dlt)_0 &=& C(\QQ), \nle
(\im \dlt)_0 &=& C(\QQ)\Ea \equiv \N.
\eens
Thus $H_\cl^0(\dlt) = C(\QQ)/\N = C(\Sigma)$ is identified with the
space of functions over physical phase space $\Sigma$.
Since we assume that there are no non-trivial relations among the
$\Ea$, the higher cohomology groups vanish. This is a standard
result \cite{HT92}. The complex (\ref{complex1}) thus gives us
a resolution of $C(\Sigma)$, which by definition means that
$H_\cl^0(\dlt) = C(\Sigma)$, $H_\cl^k(\dlt) = 0$, for all $k>0$.

A key observation in \cite{Lar04} was that the same space $C(\Sigma)$
admits a different resolution, under the technical assumption that the
Hessian (the second functional-derivative matrix of the action) is 
invertible. Introduce canonical momenta $\pa = \dd{\fa}$ and 
$\psa = \dd{\fsa}$ for both the fields and antifields. The momenta
satisfy by definition the graded canonical commutation relations
\bes
[\pb,\fa] = \dlt^\al_\bt, &\qquad&
[\fa,\fb] = [\pa,\pb] = 0, 
\nlb{ccr*}
\{\psb,\fsa\} = \dlt_\al^\bt, &\qquad&
\{\fsa,\fsb\} = \{\psa,\psb\} = 0.
\eens
We denote by $\PP$ the phase space of arbitrary histories, 
with basis $(\fa,\pb)$,
and by $\PP^*$ the extended phase space with basis
$(\fa,\pb,\ab\fsa,\psb)$. 

The definition of the KT differential extends to $\PP^*$ by requiring that
$\dlt F = [Q_{KT},F]$ for every $F \in C(\PP^*)$, where the KT operator is
$Q_{KT} = \Ea \psa$. Let us prove that this formula indeed yields a
nilpotent differential $\dlt$ defined on all of $C(\PP^*)$. 
For brevity, set $\phi^A=(\fa,\fsa)$ and $\pi_A = (\pa,\psa)$.
For any functional $F(\phi)$, 
\bes
\dlt^2 [\pi_A, F] &=& \dlt([\dlt\pi_A, F] \pm [\pi_A, \dlt F] \nl
&=& [\dlt^2\pi_A, F] \mp [\dlt\pi_A,\dlt F]
\pm [\dlt\pi_A, \dlt F] + [\pi_A, \dlt^2 F] \nle
&=& [\dlt^2\pi_A, F] \nl
&=& \dlt^2 (\d_A F) = 0,
\eens
because $\d_A F$ is a functional of $\phi$. Hence we conclude that
$[\dlt^2\pi_A, F(\phi)] = 0$ for every $F$. On the other hand, $Q_{KT}$
is linear in $\pi_A$, so there must exist some functions $f^B_A(\phi)$
such that $\dlt^2\pi_A = f^B_A(\phi)\pi_B$. But since
$f^B_A(\phi)\d_B F(\phi) \equiv 0$ for all $F$, we conclude that
$f^B_A$ must vanish themselves. Hence $\dlt^2\pi_A = 0$, and the 
definition of $\dlt$ extends to the momenta. QED.

Because of this observation, we will usually not write down the action
of the KT differential on the momenta, but only on the fields and 
antifields. Moreover, the same result holds for any nilpotent operator
which is linear in momenta, such as the longitudinal and BRST differential
encountered below.

The new ingredient in the present paper is that we assume that there 
are some relations between the EL equations (\ref{EL}). In other words,
let there be identities of the form
\be
r^\al_a\Ea \equiv 0,
\label{Rident}
\ee
where the $r^\al_a$ are some functionals of $\fa$. 
The zeroth cohomology group  $H_\cl^0(\dlt) = C(\QQ)/\N = C(\Sigma)$
is not changed, but the higher cohomology groups no longer vanish, since
$\dlt(r^\al_a\fsa) = r^\al_a\Ea \equiv 0$. The standard method to kill 
this unwanted cohomology is to introduce a bosonic second-order 
antifield $\za$, so that $r^\al_a\fsa = \dlt\za$ is KT exact.
The differential (\ref{dlt0}) is thus modified to read
\bes
\dlt \fa &=& 0, \nl
\dlt \fsa &=& \Ea, 
\label{dlt1}\\
\dlt \za &=& r^\al_a\fsa.
\eens
By introducing canonical momenta $\chi^a = \dd{\za}$ for the second-order
antifields, we can write the KT differential as a bracket, $\dlt F =
[Q_{KT},F]$, where the full KT operator is
\be
Q_{KT} =  \Ea\psa + r^\al_a\fsa\chi^a.
\label{KT}
\ee
$Q_{KT}$ is an operator in the extended phase space $\PP^*$ 
with basis $(\fa,\pb,\ab\fsa,\psb,\ab\za,\chi^b)$, and 
$\{Q_{KT}, Q_{KT}\} = 0$.

The identity (\ref{Rident})
implies that $J_a = r^\al_a\pa$ generate a Lie algebra under the Poisson
bracket. Namely, all $J_a$'s preserve the action, because
\be
[J_a, S] = r^\al_a[\pa,S] = r^\al_a\Ea \equiv 0,
\ee
and the bracket of two operators which preserve some structure also 
preserves the same structure.
We will only consider the case that the $J_a$'s generate a proper Lie
algebra $\oj$ with structure constants $f_{ab}{}^c$,
\be
[J_a, J_b] = f_{ab}{}^c J_c.
\label{oj}
\ee
The formalism extends without too much extra work to the more general
case of structure functions $f_{ab}{}^c(\phi)$, but we will not need
this complication in this paper.
It follows that the functions $r^\al_a$ satisfy the identity
\be
\db r^\al_b r^\bt_a - \db r^\al_a r^\bt_b = f_{ab}{}^c r^\al_c.
\label{rident}
\ee
The Lie algebra $\oj$ also acts on the antifields:
\bes
[J_a, \fa] &=& r^\al_a, \nl
{[}J_a, \fsa] &=& -\da r^\bt_a \fsb
\label{Jf}\\
{[}J_a, \zb] &=& f_{ab}{}^c\zc.
\eens
In particular, it follows that $\fsa$ carries a $\oj$ representation 
because it transforms in the same way as $\pa$ does.

Classically, it is always possible to reduce the phase space further,
by identifying points on $\oj$ orbits. To implement this additional
reduction, we introduce ghosts $c^a$ with anti-field number
$\afn c^a = -1$, and ghost momenta $b_a$ satisfying
\be
\{b_a, c^b\} = \dlt^b_a.
\ee
The Lie algebra $\oj$ acts on the ghosts as
\be
[J_a, c^b] = -f_{ac}{}^b c^c.
\ee
The full extended phase space, still denoted by $\PP^*$, is spanned by
$(\fa,\pb,\ab \fsa,\psb,\ab \za,\chi^b,\ab c^a,b_b)$.
The generators of $\oj$ are thus identified with the following vector
fields in $\PP^*$:
\bes
J_a &=& r^\al_a\pa -\da r^\bt_a\fsb\psa + f_{ab}{}^c\zc\chi^b
- f_{ab}{}^c c^b b_c 
\nlb{Ja}
&=& J^{field}_a + J^{ghost}_a,
\eens
where $J^{ghost}_a = f_{ab}{}^c c^b b_c$ and $J^{field}_a$ is the rest.

Now define the longitudinal derivative $d$ by
\bes
d c^a &=& -\half f_{bc}{}^a c^b c^c, \nl
d \fa &=& r^\al_a c^a, \nle
d \fsa &=& \da r^\bt_a \fsb c^a, \nl
d \za &=& - f_{ab}{}^c \zc c^b.
\eens
The longitudinal derivative can be written as $dF = [Q_{Long},F]$
for every $F\in C(\QQ^*)$, where 
\be
Q_{Long} = J^{field}_a c^a - \half f_{ab}{}^c c^a c^b b_c 
= J^{field}_a c^a + \half J^{ghost}_a c^a.
\label{QLong1}
\ee

One verifies that $d^2 = 0$ when acting on the fields and antifields by
means of the identify (\ref{rident}) and the Jacobi identities for $\oj$.
Moreover, it is straightforward to show that $d$ anticommutes with the
KT differential, $d\dlt = -\dlt d$; the proof is again done by checking
the action on the fields.
Hence we may define the nilpotent {\em BRST derivative} $s = \dlt+d$,
\bes
s c^a &=& -\half f_{bc}{}^a c^b c^c, \nl
s \fa &=& r^\al_a c^a, 
\nlb{sfields}
s \fsa &=& \Ea + \da r^\bt_a \fsb c^a, \nl
s \za &=& r^\al_a\fsa - f_{ab}{}^c \zc c^b.
\eens
Nilpotency immediately follows because 
$s^2 = \dlt^2 + \dlt d + d\dlt + d^2 = 0$.
The BRST operator can be written in the form $sF = [Q_{BRST},F]$ with
\bes
Q_{BRST} &=& Q_{KT} + Q_{Long} \nl
&=&  \Ea\psa + r^\al_a\fsa\chi^a + J^{field}_a c^a 
+ \half J^{ghost}_a c^a \nle
&=&  -\half f_{ab}{}^c c^a c^b b_c + r^\al_a c^a \pa
+ (\Ea + \da r^\bt_a \fsb c^a) \psa \nl
&&+ (r^\al_a\fsa - f_{ab}{}^c \zc c^b) \chi^a.
\eens

Like $C(\QQ)$, the function space $C(\PP^*)$
decomposes into subspaces of fixed antifield number,
$C(\PP^*) = \sum_{k=-\infty}^\infty C^k(\PP^*)$.
We can therefore define a BRST complex in $C(\PP^*)$
\be
\ldots \larroww{s} C^{-2} \larroww{s} 
C^{-1} \larroww{s} 
C^0 \larroww{s} C^1 \larroww{s} 
C^2 \larroww{s} \ldots
\label{complex2}
\ee
It is important that the spaces $C^k$ in (\ref{complex2}) are phase
spaces, equipped with the Poisson bracket (\ref{ccr*}). Unlike the
resolution (\ref{complex1}), the new resolution (\ref{complex2})
therefore allows us to do canonical quantization already in $C(\PP^*)$:
replace Poisson brackets by commutators and represent the graded
Heisenberg algebra on a Hilbert space. To pick the correct
Hilbert space, we must define a Hamiltonian which is bounded on below.
Note that different choices may be inequivalent, because there is
no Stone-von Neumann theorem in infinite dimension.

In non-covariant quantization, we single out a privileged variable $t$
among the $\al$'s, and declare it to be time.
We thus make the substitution
$\fa \to \fa(t)$, $\fsa \to \fsa(t)$, $\za \to \za(t)$, $c^a \to c^a(t)$,
$\Ea \to \Ea(t)$, 
and similar for the momenta.
The constraints (\ref{Rident}) take the form
\be
\int dt'\ r^\al_a(t,t')\Ea(t') \equiv 0
\label{rEt}
\ee
and the Lie algebra (\ref{oj}) becomes
\be
[J_a(t), J_b(t')] = \int dt''\ f_{ab}{}^c(t,t',t'') J_c(t'').
\label{ojt}
\ee
The BRST action on the fields (\ref{sfields}) is replaced by 
\bes
s c^a(t) &=& -\half \iint dt'dt''\ 
f_{bc}{}^a(t,t',t'') c^b(t') c^c(t''), \nl
s \fa(t) &=& \int dt'\ r^\al_a(t,t') c^a(t'), 
\nlb{BRST1}
s \fsa(t) &=& \Ea(t) + \iint dt'dt''\ 
\da r^\bt_a(t,t',t'') \fsb(t') c^a(t''), \nl
s \za(t) &=& \int dt'\ r^\al_a(t,t')\fsa(t') 
- \iint dt'dt''\  f_{ab}{}^c(t,t',t'') \zc(t') c^b(t''),
\eens
where we defined
\be
\da r^\bt_a(t,t',t'') \equiv {\dlt r^\bt_a(t,t')\/\dlt\fa(t'')}.
\ee
The Hamiltonian in history phase space is the generator of rigid
time translations, {\em viz.}
\be
H = -i\int dt\ \Big( \dot\fa(t)\pa(t) + \dot\fsa(t)\psa(t)
+ \dot\za(t)\chi^a(t) + \dot c^a(t)b_a(t) \Big).
\label{Hamc}
\ee
The Hamiltonian acts on the fields as
\bes
[H, \fa(t)] = -i\dot\fa(t), &\qquad&
[H, \fsa(t)] = -i\dot\fsa(t), \nle
{[}H, \za(t)] = -i\dot\za(t), &\qquad&
[H, c^a(t)] = -i\dot c^a(t).
\eens
The action on the momenta follows similarly from (\ref{Hamc}).

Expand all fields in a Fourier integral with respect to time, e.g,
\be
\fa(t) = \int_{-\infty}^\infty dm\ \fa(m) \e^{imt}.
\ee
The Hamiltonian acts on the Fourier modes as
\bes
[H, \fa(m)] = m\fa(m), &\qquad&
[H, \fsa(m)] = m\fsa(m), \nle
{[}H, \za(m)] = m\za(m), &\qquad&
[H, c^a(m)] = m c^a(m).
\eens

Now quantize. In the spirit of BRST quantization, our strategy is to
quantize first and impose dyna\-mics afterwards.
In the extended history phase space $\PP^*$, we
define a Fock vacuum $\ket 0$ which is annihilated by all negative
frequency modes, i.e.
\bes
&&\fa(-m)\ket 0 = \fsa(-m)\ket 0 = \za(-m)\ket 0 
=  c^a(-m)\ket 0 = 0, \nle
&&\pa(-m)\ket 0 = \psa(-m)\ket 0 = \chi^a(-m)\ket 0 
=  b_a(-m)\ket 0 = 0,
\eens
for all $-m < 0$. We must also decide which of the zero modes that
annihilate the vacuum, but this decision will
not affect the eigenvalues of the Hamiltonian.

The Hamiltonian (\ref{Hamc}) does not act in a well-defined manner, 
because it assigns an infinite energy to the Fock vacuum. To correct
for that, we replace the Hamiltonian by
\bes
H &=& -i\int dt\ \Big( \no{\dot\fa(t)\pa(t)} + \no{\dot\fsa(t)\psa(t)}
\nlb{Hamq}
&&+ \no{\dot\za(t)\chi^a(t)} + \no{\dot c^a(t)b_a(t)} \Big),
\eens
where normal ordering $\no{\cdot}$ moves negative frequency modes to
the right and positive frequency modes to the left. The vacuum has 
zero energy as measured by the normal-ordered Hamiltonian, $H\ket0 = 0$.
The history Hilbert space $\HH(\PP^*)$ can be identified with the
space of functions of $\fa(m)$, $\fsa(m)$, $\za(m)$, $ c^a(m)$, and
$\pa(m)$, $\psa(m)$, $\chi^a(m)$, $ b_a(m)$, where all $m > 0$.
The energy of a state $\phi^{a_1}(m_1)...b_{a_n}(m_n)\ket 0$
in $\HH(\PP^*)$ is simply the total frequency $(m_1 + ... + m_n)$.

In the absense of gauge symmetries, the BRST operator reduces to the
KT operator (\ref{KT}), which is already normal ordered. 
Writing out the time dependence explicity, we have
\bes
Q_{KT} &=& \int dt\ \no{\Ea(t)\psa(t)} 
+ \iint dtdt'\ \no{ r^\al_a(t,t')\fsa(t)\chi^a(t')} \nl
&=& \int dt\ \Ea(t)\psa(t)
+ \iint dtdt'\ r^\al_a(t,t')\fsa(t)\chi^a(t'),
\label{QKT2}
\ees
since $\Ea(t)$ and $ r^\al_a(t,t')$ depend on the fields $\fa(t)$ only,
and not on the antifields $\fsa(t)$ and $\za(t)$. This means that
quantum effects do not spoil the nilpotency of $Q_{KT}$, so the 
KT complex is well defined on the quantum level. However, the BRST
operator is not necessarily nilpotent.
The dangerous part is the longitudinal operator
\be
Q_{Long} = \int dt\ \Big( \no{J^{field}_a(t)} c^a(t) 
+ \half \no{ J^{ghost}_a(t) c^a(t) } \Big).
\label{QLong2}
\ee
The longitudinal operator ceases to be nilpotent unless the 
normal-ordered gauge generators
$J_a(t) = \no{J^{field}_a(t)} + \no{ J^{ghost}_a(t)}$
generate the algebra (\ref{oj}) without additional quantum corrections.
If so, the BRST operator also ceases to be nilpotent.

However, everything is not lost. The KT operator is still nilpotent, and
we can implement dyna\-mics as the KT cohomology in the extended phase space
without ghosts. The physical phase space now grows, because some gauge
degrees of freedom become physical upon quantization. The classical gauge
symmetry has now become an ordinary, ``global'' symmetry (albeit still
local in spacetime), which must be realized as well-defined, unitary
operators. This is a highly non-trivial requirement, which typically fails
in interesting situations. We have nothing to say anything about unitary,
but even the lesser condition that the gauge generators are at all
operators typically fails for local symmetries.

Consider for concreteness the case of Yang-Mills theory, where the gauge
generators satisfy the current algebra 
$[J_a(m), J_b(n)] = f_{ab}{}^c J_c(m+n)$, with notation as in 
(\ref{affine}). The algebra typically acts linearly on the Fourier
components of matter fields, i.e. as 
$[J_a(m), \phi(n)] = T_a\phi(m+n)$, where the $T_a$ are representation
matrices.
If the time component of $m$ is $m_0$, the normal-ordered generators read
\bes
J_a(m) &=& \sum_{n\in\ZZ^N} \no{\pi(n)T_a\phi(m-n)} 
\label{noJa}\\
&=& \sum_{n_0<m_0/2} \pi(n)T_a\phi(m-n) 
+ \sum_{n_0>m_0/2} T_a\phi(m-n)\pi(n).
\eens
They satisfy the algebra
\be
[J_a(m), J_b(n)] = f_{ab}{}^c J_c(m+n) 
+ \tr(T_aT_b)\ \dlt_{m+n}\sum_{0\leq n_0<m_0} 1.
\ee
In one dimension, the last sum is $\sum_{n_0=0}^{m_0-1} 1 = m_0$, and 
the algebra is recognized as an affine Kac-Moody algebra.
In more dimensions, however, the sum diverges. More explicitly, it 
becomes
\be
\sum_{0\leq n_0<m_0} 1 = \sum_{n_0=0}^{m_0-1} 1
\cdot \sum_{n_1=-\infty}^\infty 1 \cdot ...
= m_0 \cdot \infty^{N-1},
\ee
where $\infty$ in the last line stands for the number of integers. 
The appearence of an infinite central extension is of course nonsense,
and it shows that the normal-ordered $J_a(m)$'s are not operators.

\section{Covariant quantization}
\label{sec:covar}

The next step in \cite{Lar04} was to introduce {\em the observer's
trajectory} $q^\mu(t)$, expand all fields in a Taylor series around it, and
quantize in the space of Taylor data histories. The motivation was mainly
aesthetic; by adding the observer's trajectory, it is possible to write
down a covariant expression for the Hamiltonian, namely as the operator
which translates the fields relative to the observer. However, it is in
the presence of gauge symmetries that this construction becomes
indispensable. 

As we saw in the previous section, not only do quantum effects ruin
nilpotency of the BRST operator, but they make the gauge generators ill
defined. However, it is possible to regularize the theory formulated in
terms of Taylor data, in such a way that the full gauge symmetry of the
original model is preserved, and the regularized gauge generators are
well-defined operators. The price to pay is the appearance of an anomaly.

In the first step, we pass to a parametrized theory, and add a
parameter $t$ to each field. In other words, we make the replacements
$\fa\to\fa(t)$, $\fsa\to\fsa(t)$, etc. Contrary to the discussion in
the previous section, this $t$ is not included among the original
$\al$'s. Hence $t$ is an unphysical parameter which must eventually
be removed. In order to implement
$t$-independence in cohomology, we add the corresponding
differential. To keep the notation compact, denote collectively by
$\fA = (\fa,\fsa,\za,c^a)$ the set of all fields and antifields. We
assume that there is a nilpotent differential $\dlt$, which may be KT
or BRST,
such that
\be
\dlt\fA = z^A(\phi).
\ee
Nilpotency leads to the condition
\be
\dlt^2 \fA = z^B \d_B z^A \equiv 0.
\ee
Now replace $\fA\to\fA(t)$ and introduce addional antifields and
antighosts $\wfA(t)$, whose job is to cancel $t$-dependence in cohomology.
Define two differentials $\dlt$ and $\sigma$ by
\bes
\dlt\fA(t) &=& z^A(t), \nl
\dlt\wfA(t) &=& \wfB(t) \d_B z^A(t), 
\nlb{dltsi}
\sigma\fA(t) &=& 0, \nl
\sigma\wfA(t) &=&\d_t\fA(t),
\eens
where $\d_t = \d/\d t$ is the $t$ derivative. 
One verifies that $\dlt^2\wfA(t) = 0$, that $\sigma^2=0$ and that
$\dlt\sigma = -\sigma\dlt$. Since $\dlt^2\fA(t) = 0$ by assumption, 
the combined differential $\dlt+\sigma$ is
nilpotent. The $\sigma$ cohomology is readily computed. $H^0(\sigma)$
is spanned by functionals $\fA(t)$ which satisfy $\d_t\fA(t) = 0$, i.e.
$t$-independent functions. Hence $H^0(\dlt+\sigma) = H^0(\dlt)$. This
result is of course not unexpected. Nothing has neither been gained nor
lost by first adding a parameter $t$ and then immediately removing it in
cohomology. The reason for this exercise is rather that $t$-dependent
fields arise from Taylor expansions, and whereas physical fields must be
independent of $t$.

We are really interested in field theories, where the fields and
antifield $\fA(x)$ depend on the spacetime coordinate $x\in\RR^N$.
The parametrized fields $\fA(x,t)$ can be expanded in a Taylor
series around the observer's trajectory $q^\mu(t) \in \RR^N$:
\be
\fA(x,t) = \sum_\mm {1\/\mm!} \fAm(t)(x-q(t))^\mm,
\label{Taylor}
\ee
where $\mm = (m_1, \ab m_2, \ab ..., \ab m_N)$, all $m_\mu\geq0$, is a 
multi-index of length $|\mm| = \sum_{\mu=1}^N m_\mu$,
$\mm! = m_1!m_2!...m_N!$, and
\be
(x-q(t))^\mm = (x^1-q^1(t))^{m_1} (x^2-q^2(t))^{m_2} ...
 (x^N-q^N(t))^{m_N}.
\label{power}
\ee
Denote by $\mu$ a unit vector in the $\mu$:th direction, so that
$\mm+\mu = (m_1, \ab ...,m_\mu+1, \ab ..., \ab m_N)$. The Taylor
coefficient
\be
\fAm(t) = \d_\mm\fA(q(t),t)
= \underbrace{\d_1 .. \d_1}_{m_1} .. 
\underbrace{\d_N .. \d_N}_{m_N} \fA(q(t),t)
\label{jetdef}
\ee
is recognized as the $|\mm|$:th order mixed partial derivative of
$\fA(x,t)$, evaluated on the observer's trajectory $q^\mu(t)$.

The Taylor coefficients $\fAm(t)$ are referred to as {\em jets}; more
precisely, infinite jets. Similarly, we define a {\em $p$-jet} by
truncation to $|\mm|\leq p$; this will play an important role as a
regularization of the symmetry generators. Expand also the
Euler-Lagrange equations and the constraints in a similar Taylor
series,
\bes
\Ea(x,t) &=& \sum_\mm {1\/\mm!} \Eam(t)(x-q(t))^\mm, \nle
r^\al_a(x,t) &=& \sum_\mm {1\/\mm!} r^\al_{a\cmm}(t)(x-q(t))^\mm,
\eens
etc. These relations define the jets $\Eam(t)$ and $r^\al_{a\cmm}(t)$.
Given two jets $f_\cmm(t)$ and $g_\cmm(t')$, we define their product
\be
(f(t)g(t'))_\cmm = \sum_\nn {\mm\choose\nn}f_\cnn(t) g_{,\mm-\nn}(t').
\label{product}
\ee
It is clear that $(f(t)g(t'))_\cmm$ is the jet corresponding to the
field $f(x,t)g(x,t')$. For brevity, we also denote
$(fg)_\cmm(t) = (f(t)g(t))_\cmm$.

The equations of motion, the constraints, and the time-independence
condition translate into
\bes
\Eam(t) &=& 0, \nl
\int dt'\ (r^\al_a(t,t')\Ea(t'))_\cmm &=& 0, 
\label{mconds}\\
D_t\fam(t) \equiv \dot\fam(t) 
- \sum_\mu \dot q^\mu(t)\fa_{\cmm+\mu}(t) &=& 0.
\eens
The BRST differential $s$ which implements these conditions is
obtained from (\ref{BRST1}) by Taylor expansion:
\bes
s c^a_\cmm(t) &=& -\half \iint dt'dt''\ 
(f_{bc}{}^a(t,t',t'') c^b(t') c^c(t''))_\cmm, \nl
s \fam(t) &=& \int dt'\ (r^\al_a(t,t') c^a(t'))_\cmm, 
\label{sfields3}\\
s \fsam(t) &=& \Eam(t) + \iint dt'dt'' 
(\da r^\bt_a(t,t',t'') \fsb(t') c^a(t''))_\cmm, \nl
s \zam(t) &=& \int dt'\ (r^\al_a(t,t')\fsa(t'))_\cmm \nl
&&- \iint dt'dt''\ (f_{ab}{}^c(t,t',t'')\zc(t') c^b(t''))_\cmm.
\eens
The classical cohomology group $H_\cl^0(s)$ consists of linear 
combinations of jets $\fam(t)$ satisfying the equations (\ref{mconds})
modulo gauge transformations.

In \cite{Lar04}, we also discussed the equation of motion for
the observer's trajectory. In Minkowski spacetime, it is natural to 
assume that $q^\mu(t)$ satisfies the geodesic equation
$\ddot q^\mu(t) = 0$, which gives a contribution to the BRST
differential
\be
s\qsmu(t) = \eta_{\mu\nu} \ddot q^\nu(t).
\ee
$H_\cl^0(s)$ only contains trajectories which are straight lines,
\be
 q^\mu(t) = u^\mu t + a^\mu,
\label{line}
\ee
where $u^\mu$ and $a^\mu$ are constant vectors and $u_\mu u^\mu = 1$. 
This condition fixes
the scale of the parameter $t$ in terms of the Minkowski metric, so
we may regard it as proper time rather than as an arbitrary parameter.

Now introduce the canonical jet momenta $\pAm(t)$,
and momenta $p_\mu(t)$ and $\psmu(t)$ for the observer's trajectory 
and its antifield. The non-zero brackets are
\bes
[\pAm(t), \fBn(t')] &=& \dlt_A^B \dlt^\mm_\nn \dlt(t-t'), \nl
{[} p_\mu(t),  q^\nu(t')] &=& \dlt_\mu^\nu \dlt(t-t'), \\
{[}\psmu(t), \qsnu(t')] &=& \dlt^\mu_\nu \dlt(t-t').
\eens
As in \cite{Lar04}, we now define a
genuine Hamiltonian $H$, which translates the fields relative to the
observer or vice versa. Since the formulas are shortest when $H$ acts
on the trajectory but not on the jets, we make that choice, and define
\be
H = i\int dt\ (\dot q^\mu(t) p_\mu(t) + \dot\qsmu(t)\psmu(t)).
\label{H2}
\ee
Note the sign; moving the fields forward in $t$ is equivalent to moving
the observer backwards.
This Hamiltonian acts on the jets as
\bes
&&[H,  q^\mu(t)] = i\dot q^\mu(t), \nl
&&{[}H, \qsmu(t)] = i\dot\qsmu(t), \\
&&{[}H, \fAm(t)] = [H, \wfAm(t)] = 0.
\eens
Substituting this formula into (\ref{Taylor}), we get the energy of
the fields from
\be
[H, \fA(x,t)] = -i\dot q^\mu(t)\dmu\fA(x,t).
\label{Hcl}
\ee
In Minkowski space, the trajectory is the straight line (\ref{line}),
and $\dot q^\mu(t) = u^\mu$. If we take $u^\mu$ to be the constant
four-vector $u^\mu = (1,0,0,0)$, then (\ref{Hcl}) reduces to
\be
[H, \fA(x,t)] = -i {\d\/\d x^0}\fA(x,t).
\ee
The Hamiltonian (\ref{H2}) is thus a truly covariant generalization of
the energy operator.

Now we quantize the theory.
Since all operators depend on the parameter $t$, we can define
the Fourier components, e.g. 
\bes
\fAm(t) &=& \intdm \fAm(m)\e^{imt}, \nle
 q^\mu(t) &=& \intdm  q^\mu(m)\e^{imt}.
\eens
The Fock vacuum $\ket 0$ is defined to be annihilated by
all negative frequency modes, i.e.
\bes
&&\fAm(-m)\ket 0 = \pAm(-m)\ket 0 
= 0, 
\label{LER}\\
&& q^\mu(-m)\ket 0 = \qsmu(-m)\ket 0 
=  p_\mu(-m)\ket 0 = \psmu(-m)\ket 0 = 0,
\eens
for all $-m < 0$.
The normal-ordered form of the Hamiltonian (\ref{H2}) reads, in
Fourier space,
\be
H = -\intdm m( \no{ q^\mu(m) p_\mu(-m)} + \no{\qsmu(m) p_\mu(-m)} ),
\label{Hq}
\ee
where double dots indicate normal ordering with respect to frequency.
This ensures that $H\ket 0 = 0$.

Classically, the kinematical phase space is identified with the
KT cohomology group $H_\cl^0(\dlt)$, i.e.
the the space of fields $\fa(x)$ which solve $\Ea(x)=0$,
and trajectories $q^\mu(t) = u^\mu t+a^\mu$, where $u^2 = 1$.
In the physical phase space points on gauge orbits are identified, so
we identify it with the BRST cohomology group $H_\cl^0(s)$.
After quantization, the fields and trajectories become operators, which
act on the history Fock space. If the quantum BRST operator is 
nilpotent, we identify the physical Hilbert space with the BRST state
cohomology $H_\qm^0(Q_{BRST})$, which is the space
of functions of the positive-energy modes of the classical physical phase
space variables. However, if $\{Q_{BRST}, Q_{BRST}\} \neq 0$ we have
an anomaly, and the BRST cohomology is not well-defined on the quantum
level. Instead, we must then identify the physical Hilbert space with
the KT cohomology $H_\qm^0(Q_{KT})$; this is always well-defined
because $Q_{KT}$ is always nilpotent.

We can explicitly write down the KT operator (\ref{QKT2}) and 
longitudinal operator (\ref{QLong2}) in
jet space, and thus also $Q_{BRST} = Q_{KT}+Q_{Long}$.
If the EL equations $\Ea$ are of order $2$ and the Noether
identities $r^\al_a\Ea$ of order $3$, they read
\bes
Q_{KT} &=& 
\summ{p-2} \int dt\ \no{\Eam(t)\psam(t)} 
\label{QKT3}\\
&&+ \summ{p-3} \int dtdt'\ \no{ (r^\al_a(t,t')\fsa(t'))_\cmm
\chi^{a\cmm}(t)}, \nl
Q_{Long} &=& 
\summ{p-2} \int dtdt'\ 
\no{(r^\al_a(t,t')c^a(t'))_\cmm\pam(t)} \\
&&+ \summ{p-2} \int dtdt'dt''\ 
\no{(\da r^\bt_a(t,t',t'') \fsb(t') c^a(t''))_\cmm\psam(t)} \nl
&&- \summ{p-2} \int dtdt'dt''\ 
\no{(f_{ab}{}^c(t,t',t'')\zc(t')c^b(t''))_\cmm\chi^{a\cmm}(t)} \nl
&&-\half \summ{p-2} \int dtdt'dt''\ 
\no{(f_{ab}{}^c(t,t',t'') c^a(t')c^b(t''))_\cmm b^\cmm_c(t)}.
\eens
The condition for $Q_{Long}^2 = 0$, and thus $Q_{BRST}^2 = 0$, is
that the algebra generated by the normal-ordered gauge generators
is anomaly free. 
Necessary conditions for this are discussed in the next section.

\section{Gauge algebra}
\label{sec:Gauge}

Let us now discuss the crucial issue how the algebra of gauge
transformations is represented at the quantum level. We are dealing
with two separate questions: are the gauge generators well-defined
operators after quantization, and if so, is $Q_{BRST}^2 = 0$?
For simplicity, we assume that the constraint functions and the
structure functions have the special forms
\bes
r^\al_a(t,t') &=& r^\al_a(t)\dlt(t-t'), \nle
f_{ab}{}^c(t,t',t'') &=& f_{ab}{}^c \dlt(t-t'')\dlt(t'-t'').
\eens
The constraints (\ref{rEt}) and the Lie algebra (\ref{ojt}) then become
\bes
r^\al_a(t)\Ea(t) &\equiv& 0, 
\nlb{ojt2}
[J_a(t), J_b(t')] &=& f_{ab}{}^c J_c(t)\dlt(t-t').
\eens
These assumptions are true for the gauge algebras considered in this
paper. In the non-covariant form, $t=x^0$ is the time coordinate, 
whereas the space coordinates $x^i$ are included among the indices $a$.
In the covariant formulation in jet space, $t$ is the parameter along
the observer's trajectory, whereas $a$ contains the multi-indices $\mm$ 
labelling mixed partial derivatives. However, the assumption (\ref{ojt2})
does not hold for the diffeomorphism algebra relevant to gravity.

The gauge generator (\ref{Ja}) becomes
$J_a(t) = J^1_a(t) + J^2_a(t) + J^3_a(t) + J^4_a(t)$, where
\bes
J^1_a(t) &=& \no{r^\al_a(t)\pa(t)} \nl
J^2_a(t) &=& -\da r^\bt_a(t) \no{\fsb(t)\psa(t)}, 
\nlb{Jaquant}
J^3_a(t) &=& f_{ab}{}^c \no{\zc(t)\chi^b(t)}, \nl
J^4_a(t) &=& -f_{ab}{}^c \no{c^b(t)b_c(t)}.
\eens
As usual, the double dots denote normal ordering w.r.t. frequency.
Let us focus on the first term, which explicitly reads
\be
J^1_a(t) = r^\al_a(t)\pa^<(t) + \pa^>(t)r^\al_a(t).
\ee
where
\bes
\pa(t) &=& \intdm \pa(m)\e^{imt} \equiv \pa^<(t) + \pa^>(t), \nl
\pa^<(t) &=& \int_{-\infty}^0 dm\ \pa(m)\e^{imt}, \\
\pa^>(t) &=& \int_0^\infty dm\ \pa(m)\e^{imt}.
\eens
We also need to mode-expand the delta-function,
\bes
\dlt(t) &=& {1\/2\pi} \intdm \e^{imt} \equiv \dlt^<(t) + \dlt^>(t), \nl
\dlt^<(t) &=& {1\/2\pi} \int_{-\infty}^0 dm\ \e^{imt}, \\
\dlt^>(t) &=& {1\/2\pi} \int_0^\infty dm\ \e^{imt}.\eens

Now we further assume that
\be
[\pa(t), r^\bt_b(t')] = \da r^\bt_b(t') \dlt(t-t'),
\label{parbb}
\ee
from which it follows that
\be
[\pa^<(t), r^\bt_b(t')] = \da r^\bt_b(t') \dlt^<(t-t').
\ee
The assumption (\ref{parbb}) does hold in the gauge algebra examples
below. Using the identity
\be
\dlt^>(t-t')\dlt^<(t'-t) - \dlt^>(t'-t)\dlt^<(t-t') 
= \tpi\dot\dlt(t-t'),
\ee
it is straightforward to show that the $J^1_a$'s satisfy an extension
of the gauge algebra,
\be
[J^1_a(t), J^1_b(t')] = f_{ab}{}^c J^1_c(t)\dlt(t-t')
+ \tpi \da r^\bt_b(t') \db r^\al_a(t) \dot\dlt(t-t').
\ee
This is a well-defined expression provided that there are only finitely 
many degrees of freedom for each $t$, so the sums over $\al$ and $\bt$
are finite. This is the situation encountered in quantum-mechanical
systems, and also in field theories in one dimension. For field 
theories in several dimensions, this condition is not satisfied. Even
leaving the question of anomaly freedom aside, we need to make the
extension finite in order for $J^1_a$ to be a well-defined operator.
The way to do this is to truncate the Taylor expansion (\ref{Taylor})
at order $p$. In this way, we replace the field
$\fa(x,t)$, which has infinitely many components for each $t$ (labelled
by $x\in\RR^N$), by the $p$-jet $\fam(t)$, which has finitely many
components for each $t$ (the number of multi-indices with $|\mm|\leq p$
equals ${N+p\choose N}$).

If we write the extended algebra as
\be
[J_a(t), J_b(t')] = f_{ab}{}^c J_c(t-t') + \ext(J_a(t), J_b(t')),
\ee
and we observe that fermions give opposite signs,
we find that the extensions are
\bes
\ext(J^1_a(t), J^1_b(t')) &=&  
\tpi \da r^\bt_b(t') \db r^\al_a(t) \dot\dlt(t-t'), \nl
\ext(J^2_a(t), J^2_b(t')) &=&  
-\tpi \da r^\bt_b(t')) \db r^\al_a(t) \dot\dlt(t-t'), 
\nlb{Jaext1}
\ext(J^3_a(t), J^3_b(t')) &=& 
\tpi f_{ac}{}^d  f_{bd}{}^c \dot\dlt(t-t'), \nl
\ext(J^4_a(t), J^4_b(t')) &=& 
-\tpi f_{ac}{}^d  f_{bd}{}^c \dot\dlt(t-t').
\eens

Na\"\i vely, $\ext(J^1_a(t), J^1_b(t')) = -\ext(J^2_a(t), J^2_b(t'))$
and $\ext(J^3_a(t), J^3_b(t')) = -\ext(J^4_a(t), J^4_b(t'))$, so it 
would seem that the total extension cancels. However, a more careful 
treatment in jet space shows that this is not the case.
If $\fa(x)$ is a bosonic field, the corresponding jet $\fam(t)$ is
defined for $|\mm|\leq p$. The EL equations are second order and the
Noether identities $r^\al_a\Ea$ are third order, so the antifield
$\fsam$ is defined for $|\mm|\leq p-2$, and the second-order antifield
$\zam$ for $|\mm|\leq p-3$. Finally, the ghost $c^a(x)$ must typically
be defined for $|\mm|\leq p+1$. Hence (\ref{Jaquant}) should be written
more carefully as
\bes
J^1_a(t) &=& \summ{p} \no{r^\al_{a\cmm}(t)\pam(t)} \nl
J^2_a(t) &=& -\summ{p-2} \no{(\da r^\bt_a\fsb)_\cmm(t)\psam(t)}, 
\nlb{Jaquant2}
J^3_a(t) &=& f_{ab}{}^c \summ{p-3} \no{\zcm(t)\chi^{b\cmm}(t)}, \nl
J^4_a(t) &=& -f_{ab}{}^c \summ{p+1} \no{c^b_\cmm(t)b_c^\cmm(t)}.
\eens
In analogy with (\ref{parbb}), we assume that
\be
[\pam(t), r^\bt_{b\cnn}(t')] = \dam r^\bt_{b\cnn}(t') \dlt(t-t').
\ee
The jet form of the extensions (\ref{Jaext1}) read
\bes
\ext(J^1_a(t), J^1_b(t')) &=&  
\tpi \summ{p}\sumn{p}
\dam r^\bt_{b\cnn}(t') \dbn r^\al_{a\cmm}(t) \dot\dlt(t-t'), \nl
\ext(J^2_a(t), J^2_b(t')) &=&  
-\tpi \summ{p-2}\sumn{p-2}
\dam r^\bt_{b\cnn}(t') \dbn r^\al_{a\cmm}(t) \dot\dlt(t-t'), \nl
\nlb{Jaext2}
\ext(J^3_a(t), J^3_b(t')) &=&  
\tpi f_{ac}{}^d  f_{bd}{}^c {N+p-3\choose N} \dot\dlt(t-t'), \nl
\ext(J^4_a(t), J^4_b(t')) &=&  
-\tpi f_{ac}{}^d  f_{bd}{}^c {N+p+1\choose N} \dot\dlt(t-t').
\eens
Note in particular that $f_{ac}{}^d  f_{bd}{}^c$ is proportional to
the second Casimir operator.

The appearance of gauge anomalies is thus a generic feature of canonical
quantization of field theories in history space. In exceptional cases,
such as the free Maxwell field to be discussed next, there are no
anomalies, and we can pass to the BRST cohomology, but generically
there are anomalies and we must identify the physical Hilbert space
with the KT cohomology only. Moreover, the crucial nature of the
truncation to $p$-jets becomes clear; in (\ref{Jaext1}) the sums over
$\al$ and $\bt$ diverge (for field theories), whereas the sums over
$\mm$, $\nn$, $\al$ and $\bt$ in (\ref{Jaext2}) are finite and thus
well defined. The sums diverge in the $p\to\infty$ limit, however, 
making it difficult to remove the regularization.

\section{The free Maxwell field}
\label{sec:Maxwell}

The Maxwell field $A_\mu(x)$ transforms as a vector field under the 
Poincar\'e group and as a connection under the gauge algebra 
$\map(N, \uu(1))$, whose smeared generators
are denoted by $\J_X = \int \dNx X(x) J(x)$:
\be
[\J_X, A_\mu(x)] = \dmu X(x).
\label{u1}
\ee
We use the Minkowski metric $\eta_{\mu\nu}$ and its inverse 
$\eta^{\mu\nu}$ to freely raise and lower indices, e.g.
$F^{\mu\nu} = \eta^{\mu\rho}\eta^{\nu\si}F_{\rho\si}$.
The field strength $F_{\mu\nu}(x) = \dmu A_\nu(x) - \dnu A_\mu(x)$
transforms in the adjoint representation, i.e. trivially.
The action 
\be
S = \quart \int \dNx F_{\mu\nu}(x)F^{\mu\nu}(x)
\ee
leads to the equations of motion
\be
\EE^\mu(x) \equiv -{\dlt S\/\dlt A_\mu(x)} 
= \dnu F^{\mu\nu}(x) = 0.
\label{EMax}
\ee
The Maxwell equations are not all independent, because of the 
constraints
\be
\dmu \EE^\mu(x) = \dmu\dnu F^{\mu\nu}(x) \equiv 0.
\label{cMax}
\ee
We are thus instructed to introduce
the following fields: the first-order antifield $\As^\mu(x)$
for the EL equation $\dnu F^{\mu\nu}(x) = 0$; the second-order antifield
$\zeta(x)$ for the identity $\dmu\dnu F^{\mu\nu}(x) \equiv 0$;
and the ghost $c(x)$ to identify fields related by a gauge transformation
of the form (\ref{u1}).

The BRST operator $s$ acts as
\bes
s c(x) &=& 0, \nl
s A_\mu(x) &=& \dmu c(x), 
\nlb{sMax1}
s \As^\mu(x) &=& \dnu F^{\mu\nu}(x), \nl
s \zeta(x) &=& \dmu \As^\mu(x),
\eens
We check that $s F_{\mu\nu} = s \dmu\As^\mu = 0$, so the kernel of $s$
is spanned by $c$, the field strengths $F_{\mu\nu}$, and
$\dmu\As^\mu$. $\im s$ is generated by the ideals $\dmu c$,
$\dnu F^{\mu\nu}$, and $\dmu \As^\mu$. Hence $H_\cl^\bullet(s)$ 
consists of the gauge-invariant parts of $A_\mu$ (i.e. $F_{\mu\nu}$) 
which solve the Maxwell equations, as expected.

Introduce canonical momenta $E^\mu(x)$, $\Es_\mu(x)$, $\chi(x)$ and
$b(x)$, defined by the following non-zero brackets:
\bes
[E^\mu(x), A_\nu(x')] &=& \dlt^\mu_\nu \dlt(x-x'), \nl
\{\Es_\mu(x), \As^\nu(x')\} &=& \dlt_\mu^\nu \dlt(x-x'), \nle
{[}\chi(x), \zeta(x')] &=& \dlt(x-x'), \nl
\{b(x), c(x')\} &=&  \dlt(x-x').
\eens
It should be emphasized that $E^\mu = \dlt/\dlt A_\mu$ is the conjugate
of the gauge potential in history space, and not yet related to the
electric field $F^{\mu0}$. We could introduce the 
condition $E^\mu = F^{\mu0}$ as a constraint in the history phase space,
turning the Maxwell equations into second class constraints. 
By keeping dyna\-mics as a first-class constraint no such condition, which
would ruin covariance, is necessary.
The BRST operator can explicitly be written as
\be
Q_{BRST} = \int \dNx \Big( \dmu c(x)E^\mu(x) 
+ \dnu F^{\mu\nu}(x)\Es_\mu(x) + \dmu \As^\mu(x)\chi(x) \Big).
\ee
{F}rom this we read off that
the action on the momenta is given by
\bes
s b(x) &=& \dmu E^\mu(x), \nl
s E^\mu(x) &=& \dmu \d^\nu \Es_\nu(x) - \dnu\d^\nu \Es_\mu(x), 
\nlb{Mxwmom}
s \Es_\mu(x) &=& \dmu\chi(x), \nl
s \chi(x) &=& 0.
\eens
Note the duality between fields and momenta; (\ref{sMax1}) and
(\ref{Mxwmom}) are identified under the replacements
\bes
b &\longleftrightarrow& \zeta, \nl
E^\mu &\longleftrightarrow& \As^\mu, \nle
E^*_\mu &\longleftrightarrow& A_\mu, \nl
\chi &\longleftrightarrow& c.
\eens

The physical content of the theory is clearer in Fourier space.
The BRST operator
\be
Q_{BRST} &=& \int \dNk \Big( k_\mu c(k)E^\mu(-k) 
+ (k^\mu k_\nu A^\nu(k) - k^\nu k_\nu A^\mu(k))\Es_\mu(-k) \nl
&&\qquad+ k_\mu \As^\mu(k)\chi(-k) \Big),
\label{BRSTMaxw}
\ee
acts on the Fourier modes as
\bes
s c(k) &=& 0, \nl
s A_\mu(k) &=& k_\mu c(k), \nle
s \As^\mu(k) &=& k^\mu k_\nu A^\nu(k) - k^\nu k_\nu A^\mu(k), \nl
s \zeta(k) &=& k_\mu \As^\mu(k).
\eens
We distinguish between two cases:

1. $k^2 = \om^2 \neq 0$, say $k = (\om,0,0,0)$. Then
$sc = 0$, $s A_0 = \om c$, $s A_1 = s A_2 = s A_3 = 0$, 
$s \As^0 = \om^2 A_0 - \om\om A_0 = 0$, $s\As^1 = \om^2 A_1$ 
$s\As^2 = \om^2 A_2$, $s\As^3 = \om^2 A_3$ and $s\zeta = \om \As^0$. 
The kernel is thus spanned by $c$, $A_1$, $A_2$, $A_3$ and $\As^0$, and
the image is spanned by $c$, $A_1$, $A_2$, $A_3$ and $\As^0$. Since 
$\ker s = \im s$ there is no cohomology.

2. $k^2 = 0$, say $k = (k_0,0,0,k_0)$. Then
$sc = 0$, $s A_0 = s A_3 = k_0 c$, $s A_1 = s A_2 = 0$,
$s \As^0 = s \As^3 = k^0 k_\nu A^\nu$, $s\As^1 = s\As^2 = 0$ and
$s\zeta = k_\mu \As^\mu$. 
The kernel is thus spanned by $c$, $A_1$, $A_2$, $k^\mu A_\mu = 
A_0 - A_3$, $\As^1$, $\As^2$, and  $k_\mu\As^\mu = \As^0 - \As^3$.
The image is spanned by $c$, $k^\mu A_\mu$ and $k_\mu\As^\mu$, which
factor out in cohomology. We are left with two physical polarizations
$A_1$ and $A_2$.

We now quantize in the history phase space before introducing dyna\-mics
by passing to the BRST cohomology.
We single out one direction $x^0$ as time, and take the Hamiltonian to 
be the generator of rigid time translations,
\bes
H &=& -i\int \dNx \Big( 
  \d_0 A_\mu(x)E^\mu(x) + \d_0 \As^\mu(x)\Es_\mu(x) \nl
&&\qquad+ \d_0 \zeta(x)\chi(x) + \d_0 c(x)b(x) \Big) \nle
&=& \int \dNk k_0 \Big( 
  A_\mu(k)E^\mu(-k) + \As^\mu(k)\Es_\mu(-k) \nl
&&\qquad+  \zeta(k)\chi(-k) + c(k)b(-k) \Big).
\eens
Note that at this stage we break Poincar\'e invariance,
since the Hamiltonian treats the $x^0$ coordinate differently from
the other $x^\mu$.
Quantize by introducing a Fock vacuum $\ket 0$ satisfying
\bes
&&A_\mu(k)\ket 0 = E^\mu(k)\ket 0 = 
\As^\mu(k)\ket 0 = \Es_\mu(k)\ket 0 = 
\nlb{LER2}
&&\zeta(k)\ket 0 = \chi(k)\ket 0 = 
c(k)\ket 0 = b(k)\ket 0 = 0,
\eens
for all $k$ such that $k_0 < 0$. 

At this point we want to pass to BRST cohomology. There might be
problems with normal ordering, but in fact the BRST operator
(\ref{BRSTMaxw}) is already normal ordered. This is because
the generator of $\uu(1)$ gauge transformations
\be
\J_X = -\int \dNx X(x)\dmu E^\mu(x)
\label{u1gauge}
\ee
is itself already normal ordered. There are thus no anomalies, and 
the BRST operator (\ref{BRSTMaxw}) remains nilpotent.
We define the BRST state cohomology as the space of physical states,
where a state is physical if it is BRST closed,
$Q_{BRST}\ket{phys} = 0$, and two physical states are equivalent if 
they differ by a BRST exact state,
$\ket{phys} \sim \ket{phys'}$ if
$\quad \ket{phys} - \ket{phys'} = Q_{BRST}\ket{}$.

The rest proceeds as for the harmonic oscillator \cite{Lar04}.
After adding a small
perturbation to make the Hessian invertible, all momenta (\ref{Mxwmom})
vanish in cohomology, and only the transverse polarizations
$\eps^\mu A_\mu(k) = 0$ with $\eps^\mu k_\mu = 0$ and $\eps^0 = 0$
survive. A basis for the history Hilbert space consists of multi-quanta 
states
\be
\eps^\mu_1A_\mu(k^{(1)})\ldots\eps^\mu_nA_\mu(k^{(n)})\ket{0}
\ee
where $k^{(j)}_\mu k^{(j)\mu} = 0$ and $k^{(j)}_0 > 0$.
The energy is given by $H = k_0^{(1)} + ...+ k_0^{(n)}$.
The gauge generators (\ref{u1gauge}) act in a well-defined manner, 
in fact trivially, on the Hilbert space, because
$\eps^\mu_jk^{(j)}_\mu = 0$.

As in Section \ref{sec:covar}, we want to give a completely covariant 
description of the Hamiltonian. To this end, we introduce the observer's
trajectory $q^\mu(t) \in \RR^N$, and expand all fields in a Taylor
series around it, i.e. we pass to jet data.
Hence e.g.,
\be
A_\mu(x) = \sum_\mm {1\/\mm!} A_{\mu\cmm}(t)(x-q(t))^\mm.
\label{STaylor}
\ee
The equations of motion (\ref{EMax}) translate into
\be
\sum_\nu F^{\mu\nu}_{\cmm+\nu}(t) = 0,
\ee
and the constraint (\ref{cMax}) becomes
\be
\sum_\mu\EE^\mu_{\mu\cmm}(t) 
= \sum_{\mu\nu} F^{\mu\nu}_{\cmm+\mu+\nu}(t)\equiv 0,
\ee
where the field strength is
\be
F_{\mu\nu\cmm}(t) = A_{\mu,\mm+\nu}(t) - A_{\nu,\mm+\mu}(t) = 0.
\ee
We introduce jets also for the antifields and for the ghost, denoted
by $A^\mu_{*,\mm}(t)$, $\zeta_{,\mm}(t)$, $c_{,\mm}(t)$. 
The BRST differential $s$ which implements all these conditions is
defined by
\bes
s c_{,\mm}(t) &=& 0, \nl
s A_{\mu,\mm}(t) &=& c_{,\mm+\mu}(t), 
\nlb{sMax}
s A^\mu_{*,\mm}(t) &=& \sum_\nu F^{\mu\nu}_{,\mm+\nu}(t), \nl
s \zeta_{,\mm}(t) &=& \sum_\mu A^\mu_{*,\mm+\mu}(t).
\eens

Moreover, we demand that the Taylor series does not depend on the
parameter $t$, which gives rise to conditions of the type
\be
D_t A_{\mu,\mm}(t) \equiv \dot A_{\mu,\mm}(t) 
- \sum_\nu \dot q^\nu(t)A_{\mu,\mm+\nu}(t) &=& 0.
\ee
As explained in (\ref{dltsi}), we need to double the number of antifields
and introduce an additional differential $\sigma$ to remove these
conditions in cohomology. Thus we introduce antifields 
$\bar c_{,\mm}(t)$, $\bar A_{\mu,\mm}(t)$, $\bar A^\mu_{*,\mm}(t)$, 
$\bar \zeta_{,\mm}(t)$ and set
\bes
\sigma \bar c_{,\mm}(t) &=& \dot c_{,\mm}(t), \nl
\sigma \bar A_{\mu,\mm}(t) &=& \dot A_{\mu,\mm}(t), \nl
\sigma \bar A^\mu_{*,\mm}(t) &=& \dot A^\mu_{*,\mm}(t), \\
\sigma \bar \zeta_{,\mm}(t) &=& \dot \zeta_{,\mm}(t) \nl
\sigma c_{,\mm}(t) &=& \sigma A_{\mu,\mm}(t) = \sigma A^\mu_{*,\mm}(t)
= \sigma \zeta_{,\mm}(t) = 0.
\eens
Clearly, $\sigma^2 = 0$.
We also extend the definition of the BRST differential $s$ to the
barred antifields:
\bes
s \bar c_{,\mm}(t) &=& 0, \nl
s \bar A_{\mu,\mm}(t) &=& -\bar c_{,\mm+\mu}(t), \nle
s \bar A^\mu_{*,\mm}(t) &=& 
-\sum_\nu (\bar A_{\nu,\mm+\mu}(t) - \bar A_{\mu,\mm+\nu}(t)), \nl
s \bar \zeta_{,\mm}(t) &=& -\sum_\mu \bar A^\mu_{*,\mm+\mu}(t).
\eens
That $s^2 = 0$ follows in the same way as for (\ref{sMax}). 
Moreover, we verify that $s\sigma = -\sigma s$, and hence $s+\sigma$
is nilpotent.

The classical cohomology group $H_\cl^0(s+\sigma)$ consists of linear 
combinations of jets satisfying
\be
A_{\mu,\mm}(t) = \eps_\mu(t) \e^{ik\cdot q(t)} (ik)^\mm
\label{fmt}
\ee
where $k^2=0$ and the polarization vector $\eps_\mu(t)$ is perpendicular
both to the photon momentum and the observer's trajectory: 
\be
\eps_\mu(t) k^\mu = \eps_\mu(t) \dot q^\mu(t) = 0.
\ee
The latter is evidently equivalent to the non-covariant condition 
$\eps^0 = 0$. Moreover, $k\cdot q = k_\mu q^\mu$ and the power
$k^\mm$ is defined in analogy with (\ref{power}).
The Taylor series (\ref{STaylor}) can be summed, giving
\bes
A_\mu(x) &=&
\e^{ik\cdot q(t)} \sum_\mm {1\/\mm!} \eps_\mu(t) (ik)^\mm(x-q(t))^\mm \nl
&=& \eps_\mu(t) \e^{ik\cdot q(t)} \e^{ik\cdot(x-q(t))}
\label{Axt}\\
&=& \eps_\mu(t) \eikx.
\eens

The physical Hamiltonian $H$ is defined as in Equation (\ref{Hq}). The
classical phase space $H_\cl^0(s+\sigma)$ is thus the the space of plane 
waves $\eikx$, cf. (\ref{Axt}),  and straight trajectories 
$q^\mu(t) = u^\mu t+a^\mu$. The energy is given by
\bes
[H, \eikx] &=& k_\mu \dot q^\mu(t)\eikx = k_\mu u^\mu \eikx, \nle
[H,  q^\mu(t)] &=& i\dot q^\mu(t) = iu^\mu.
\eens
This is a covariant description of phase space, because the energy
$k_\mu u^\mu$ is Poincar\'e invariant.

We now quantize the theory before imposing dyna\-mics. To this end,
we introduce the canonical momenta for all jets and antijets, and
$p_\mu(t)$ and $\psmu(t)$ for the observer's trajectory and its
antifield. The defining relations are
\bes
[E^{\mu,\mm}(t), A_{\nu,\nn}(t')] 
&=& \dlt^\mu_\nu \dlt^\mm_\nn \dlt(t-t'), \nl
\{E^{*\cmm}_\mu(t), A^\nu_{*\cnn}(t')\} 
&=& \dlt_\mu^\nu \dlt^\mm_\nn \dlt(t-t'), \nl
{[}\chi^\cmm(t), \zeta_\cnn(t')] &=& \dlt^\mm_\nn \dlt(t-t'), \\
\{b^\cmm(t), c_\cnn(t')\} &=&  \dlt^\mm_\nn \dlt(t-t'), \nl
{[}p_\nu(t),  q^\mu(t')] &=& \dlt^\mu_\nu \dlt(t-t'). 
\eens
Since the jets also depend on the parameter $t$, we can define
their Fourier components, e.g. 
\bes
A_{\mu,\mm}(t) &=& \intdm A_{\mu,\mm}(m)\e^{imt}, \nle
q^\mu(t) &=& \intdm  q^\mu(m)\e^{imt}.
\eens
The Fock vacuum (\ref{LER2}) is replaced by a new vacuum, also denoted 
by $\ket 0$, which is defined to be annihilated by
the negative frequency modes of the jets and antijets, e.g.
\bes
&&A_{\mu,\mm}(-m)\ket 0 = E^\mu_{,\mm}(-m)\ket 0 
= A^\mu_{*\cmm}(-m)\ket 0 
\nlb{LER3}
&&= E_\mu^{*\cmm}(-m)\ket 0 = q^\mu(-m)\ket 0 =  p_\mu(-m)\ket 0 = 0,
\eens
for all $-m < 0$.
The quantum Hamiltonian is still defined by (\ref{Hq}),
where double dots indicate normal ordering with respect to frequency,
ensuring that $H\ket 0 = 0$.

It remains to check that the algebra of $\uu(1)$ gauge transformations
acts in a well-defined manner before we can pass to the BRST cohomology.
Since a gauge potential transforms as
\be
[\J_X, A_{\mu,\mm}(t)] = \d_{\mm+\mu}X(q(t))
\ee
we have
\be
\J_X = \sum_\mm \sum_\mu \int dt\ \d_{\mm+\mu}X(q(t))E^{\mu\cmm}(t).
\label{JXMax}
\ee
There are no contributions from the antifields, since $\As^\mu$,
$\zeta$ and $c$ all transform trivially under $\map(N,\uu(1))$.
The expression (\ref{JXMax}) is evidently normal ordered as it stands, 
and  consequently there are no gauge anomalies. 

The rest proceeds as for the scalar field \cite{Lar04}.
We can consider the one-quantum state with momentum $k$ over the
true Fock vacuum, $\ket k = \exp(ik\cdot x)\ket 0$. This state is
not an energy eigenstate, because the Hamiltonian excites a quantum
of the observer's trajectory: $H\ket k = k_\mu u^\mu\ket k$.
In some approximations, we  may treat the observer's trajectory as 
a classical variable and introduce the macroscopic reference state 
$\ket{0;u,a}$, on which 
$q^\mu(t)\ket{0;u,a} = (u^\mu t + a^\mu)\ket{0;u,a}$, where $u^\mu$
and $a^\mu$ are c-numbers rather than quantum operators here.
We can then consider a state
$\ket{k,\eps; u,a} = \eps_\mu(t) \exp(ik\cdot x)\ket{0; u,a}$ with one 
quantum over the reference state. The Hamiltonian gives
$H \ket{k,\eps; u,a} = k_\mu u^\mu\ket{k,\eps;u,a}$.
In particular, if $u^\mu = (1,0,0,0)$, then
the eigenvalue of the Hamiltonian is $k_\mu u^\mu = k_0$, as
expected. Moreover, the lowest-energy condition (\ref{LER3}) ensures
that only quanta with positive energy will be excited; if 
$k_\mu u^\mu < 0$ then $\ket{k,\eps;u,a} = 0$.

\section{Fermions}
\label{sec:fermions}

In the previous section we described the free Maxwell field in covariant
canonical quantization. Apart from the explicit introduction of the
observer, which occurred already for the free scalar field, the results
were completely standard. In this section we couple the Maxwell field
to a Dirac spinor, and encounter new phenomena: the
gauge algebra acquires an extension, the BRST operator ceases to
be nilpotent, and only the KT operator is well defined.

We use standard notation; $\psi$ is a Dirac spinor and
$\bar\psi(x) = \psi^\dagger(x)\gm^0$ is the Dirac conjugate.
The Dirac action is
\be
S_D = \int \dNx \bar\psi(x)(\gm^\mu(i\dmu - eA_\mu(x)) - M)\psi(x),
\ee
where $e$ is the charge of the positron, $M$ its mass,
and $A_\mu(x)$ is the gauge potential. The Dirac equation, i.e. the
EL equation for $\psi$, reads
\bes
\gm^\mu(i\dmu-eA_\mu(x)) \psi(x) &=& M\psi(x), \nle
(i\dmu+eA_\mu(x))\bar\psi(x)\gm^\mu &=& -M\bar\psi(x).
\eens
The free Maxwell equations (\ref{EMax}) are changed into
\be
\EE^\mu(x) = \dnu F^{\mu\nu}(x) - j^\mu(x) = 0,
\label{EMax2}
\ee
where the current is
\be
j^\mu(x) = e \bar\psi(x)\gm^\mu\psi(x).
\label{jmuDir}
\ee
The EL equations are not all independent, because of the 
continuity equation
\be
\dmu \EE^\mu(x) = -\dmu j^\mu(x) \equiv 0.
\label{cMax2}
\ee

To implement the Dirac equation in cohomology, we treat $\psi(x)$
and $\bar\psi(x)$ as independent fields, and introduce the 
corresponding antifields $\psi^*(x)$ and $\bar\psi^*(x)$.
The KT differential acts on the fermions as
\bes
\dlt \psi(x) &=& 0, \nl
\dlt \bar\psi(x) &=& 0,
\nlb{KTpsi}
\dlt \psi^*(x) &=& \gm^\mu(i\dmu-eA_\mu(x)) \psi(x) - M\psi(x), \nl
\dlt \bar\psi^*(x) &=& (i\dmu+eA_\mu(x))\bar\psi(x)\gm^\mu 
+ M\bar\psi(x).
\eens
Moreover, we also need to add an additional antifield to 
implement the condition $\bar\psi(x) = \psi^\dagger(x)\gm^0$, but
we leave this implicit for brevity of notation.
The presence of the fermions changes how the KT differential acts
on the Maxwell field (\ref{sMax1}),
\bes
\dlt A_\mu(x) &=& 0, \nl
\dlt \As^\mu(x) &=& \dnu F^{\mu\nu}(x) - j^\mu(x),
\label{sMax2} \\
\dlt \zeta(x) &=& \dmu \As^\mu(x) 
- ie\bar\psi^*(x)\psi(x) - ie\bar\psi(x)\psi^*(x).
\eens
We check that
\be
\dlt^2\zeta &=& \dmu\dnu F^{\mu\nu} - \dmu j^\mu 
+ e\dmu\bar\psi\gm^\mu\phi + e\bar\psi\gm^\mu\dmu\psi \equiv 0.
\ee
The fermions transform under gauge transformations as
\bes
[\J_X,\psi(x)] &=& -eX(x)\psi(x), \nl
{[}\J_X,\bar\psi(x)] &=& eX(x)\bar\psi(x), \nle
{[}\J_X,\psi^*(x)] &=& -eX(x)\psi^*(x),\nl
{[}\J_X,\bar\psi^*(x)] &=& eX(x)\bar\psi^*(x).
\eens
The BRST differential thus gives
\bes
s \psi(x) &=& -eX(x)\psi(x)c(x), \nl
s \bar\psi(x) &=& eX(x)\bar\psi(x)c(x), 
\label{spsi}\\
s \psi^*(x) &=& \gm^\mu(i\dmu-eA_\mu(x)) \psi(x) - M\psi(x)
-eX(x)\psi^*(x)c(x), \nl
s \bar\psi^*(x) &=& (i\dmu+eA_\mu(x))\bar\psi^*(x)\gm^\mu + M\bar\psi(x)
+eX(x)\bar\psi(x)c(x).
\eens
We introduce canonical momenta $\pi(x) = \dlt/\dlt\psi(x)$,
$\bar\pi(x) = \dlt/\dlt\bar\psi(x)$, 
$\pi_*(x) = \dlt/\dlt\psi^*(x)$, and
$\bar\pi_*(x) = \dlt/\dlt\bar\psi^*(x)$, with non-zero brackets
\bes
\{\pi(x), \psi(y)\} = \dlt(x-y), &\quad&
[\pi_*(x), \psi^*(y)] = \dlt(x-y), \nle
\{\bar\pi(x), \bar\psi(y)\} = \dlt(x-y), &\quad&
[\bar\pi_*(x), \bar\psi^*(y)] = \dlt(x-y).
\eens
We can then explicitly write down the KT operator
\bes
Q_{KT} &=& \int \dNx \Big( 
(\dnu F^{\mu\nu}(x) - j^\mu(x)) \Es_\mu(x) \nl
&&+ (\dmu \As^\mu(x) 
- ie\bar\psi^*(x)\psi(x) - ie\bar\psi(x)\psi^*(x))\chi(x) \\
&&+ ( \gm^\mu(i\dmu-eA_\mu(x)) \psi(x) - M\psi(x))\pi_*(x) \nl
&&+( (i\dmu+eA_\mu(x)) \bar\psi(x)\gm^\mu + M\bar\psi(x))\bar\pi_*(x) 
\Big), 
\eens
the gauge generators,
\bes
\J_X &=& \int \dNx \Big( \dmu X(x)E(x) 
- eX(x)\psi(x)\pi(x) + eX(x)\bar\psi(x)\bar\pi(x) \nl
&&- eX(x)\psi^*(x)\pi_*(x) + eX(x)\bar\psi^*(x)\bar\pi_*(x) \Big),
\label{JXpsi}
\ees
and the BRST operator
\bes
Q_{BRST}&=& \int \dNx \Big( 
\dmu c(x)E^\mu(x) 
+ (\dnu F^{\mu\nu}(x) - j^\mu(x)) \Es_\mu(x) \nl
&&+ (\dmu \As^\mu(x)
- ie\bar\psi^*(x)\psi(x) - ie\bar\psi(x)\psi^*(x)) \chi(x) \nl
&& - eX(x)\psi(x)c(x)\pi(x) + eX(x)\bar\psi(x)c(x)\bar\pi(x) 
\label{BRSTDir}\\
&&+( \gm^\mu(i\dmu-eA_\mu(x)) \psi(x) - M\psi(x))\pi_*(x) \nl
&&+( (i\dmu+eA_\mu(x)) \bar\psi(x)\gm^\mu + M\bar\psi(x))\bar\pi_*(x) 
 \Big).
\eens

As in the previous section, we quantize first and impose dyna\-mics
afterwards. However, this requires that $Q_{BRST}$ remains a well-defined,
nilpotent operator even after normal ordering, which fails in the
presence of fermions. In fact, even the gauge generators (\ref{JXpsi})
fail to be well-defined. The dangerous part is
$\J_X = -e\int \dNx X(x)\no{\psi(x)\pi(x)}$.
With $X = \exp(ik\cdot x)$, we have
\bes
\J(k) &=& -e\int \dNx \e^{ik\cdot x}\no{\psi(x)\pi(x)} \nle
&=& -e\int \dNkp \no{\psi(k')\pi(k-k')}.
\eens
One checks that
\be
[\J(k), \J(k')] = e^2 Z(k) \dlt(k+k'),
\label{charge0}
\ee
where
\be
Z(k) = \int_0^{k_0} dk'_0 \int_{-\infty}^\infty dk'_1 \ldots
 \int_{-\infty}^\infty dk'_{N-1} = k_0 \infty^{N-1}.
\ee
The gauge algebra thus acquires a central extension. In one dimension,
this is an affine algebra, but in more dimensions the central extension
is infinite. So not only is the BRST operator not nilpotent, which we
have to live with, but the anomalous gauge generators are not
even operators, because infinities arise.

In this situation, the passage to jet space becomes indisposable, 
because it offers a way to regularize the theory so that the gauge
generators become well defined. As usual we associate a family of
jets to each field, e.g. $\psi(x)\to\psi_\cmm(t)$. The action of the
KT differential in jet space is given by
\bes
\dlt \psi_\cmm(t) &=& 0, \nl
\dlt \bar\psi_\cmm(t) &=& 0, \nl
\dlt \psi^*_\cmm(t) &=& \sum_\mu i\gm^\mu\psi_{\cmm+\mu}(t)
-e\sum_\mu \gm^\mu(A_\mu\psi)_\cmm(t) - M\psi_\cmm(t), \nl
\dlt \bar\psi^*_\cmm(t) &=& \sum_\mu i\bar\psi_{\cmm+\mu}(t)\gm^\mu
+e\sum_\mu (A_\mu\bar\psi)_\cmm(t)\gm^\mu + M\bar\psi_\cmm(t), \nl
\dlt A_{\mu\cmm}(t) &=& 0, \\
\dlt A^\mu_{*\cmm}(t) &=& 
\sum_\nu F^{\mu\nu}_{\cmm+\nu}(t) - j^\mu_\cmm(t), \nl
\dlt \zeta_\cmm(t) &=& \sum_\mu A^\mu_{*\cmm+\mu}(t)
- ie(\bar\psi^*\psi)_\cmm(t) - ie(\bar\psi\psi^*)_\cmm(t),
\eens
where
\bes
F_{\mu\nu\cmm}(t) &=& A_{\nu, \mm+\mu}(t) - A_{\mu, \mm+\nu}(t), \nle
j^\mu_\cmm(t) &=& e (\bar\psi\gm^\mu\psi)_\cmm(t).
\eens
We also introduce extra antifields to cancel the $t$ dependence, but
they will not be explicitly described here. Moreover, we introduce
jet momenta, e.g. $\pi^\cmm(t) = \dlt/\dlt\psi_\cmm(t)$.

At this point we regularize the theory; simply ignore the jets for the
original fields
with $|\mm|>p$. Since the Dirac equation is first order, Maxwell's
equations are second order, and the continuity equation is third 
order, we truncate the jets as follows
\bes
\barr{|c|c|c|l|}                                                       
\hline
\afn & \hbox{Jet} & \hbox{Momentum} & \hbox{Order} \\
\hline
0 & \psi_\cmm(t), \bar\psi_\cmm(t) 
& \pi^\cmm(t), \bar\pi^\cmm(t) & p \\
0 & A_{\mu\cmm}(t) & E^{\mu\cmm}(t) & p \\
1 & \psi^*_\cmm(t), \bar\psi^*_\cmm(t)
& \pi_*^\cmm(t), \bar\pi_*^\cmm(t)  & p-1 \\
1 & A^\mu_{*\cmm}(t) & E_\mu^{*\cmm}(t) & p-2 \\
2 & \zeta_\cmm(t) & \chi^\cmm(t) & p-3 \\
\hline
\earr
\ees
The KT operator reads explicitly
\bes
Q_{KT} &=& \int dt\ \Big\{
\summ{p-2} (\sum_{\mu\nu} F^{\mu\nu}_{\cmm+\nu}(t)E^{*\cmm}_\mu(t) 
- \sum_\mu j^\mu_\cmm(t) E^{*\cmm}_\mu(t) ) \nl
&+& \summ{p-3} (\sum_\mu A_{*\cmm+\mu}^\mu(t)
- ie(\bar\psi^*\psi)_\cmm(t) - ie(\bar\psi\psi^*)_\cmm(t)) \chi^\cmm(t) 
\nl
&+& \summ{p-1} (\sum_\mu \gm^\mu ( i\psi_{\cmm+\mu}(t) 
- e(A_\mu\psi)_\cmm(t)) - M\psi_\cmm(t))\pi_*^\cmm(t)  \nl
&+&\summ{p-1} (\sum_\mu( i\bar\psi_{\cmm+\mu}(t) + 
e(A_\mu\bar\psi)_\cmm(t))\gm^\mu + M\bar\psi_\cmm(t))\bar\pi_*^\cmm(t) 
 \Big\}. \nl
\label{QDir}
\ees
The gauge generators (\ref{JXpsi}) are truncated to
\bes
\J_X &=& \int dt\ \Big\{
\summp \sum_\mu \d_{\mm+\mu}X(q(t)) E^{\mu\cmm}(t) 
\label{JXDir}\\
&+& e\sumnmp{} \mndmn X(q(t))
(\no{\bar\psi_\cnn(t)\bar\pi^\cmm(t)}
-\no{\psi_\cnn(t)\pi^\cmm(t)})  \nl
&+& e\sumnmp{-1} \mndmn X(q(t))
(\no{\bar\psi^*_\cnn(t)\bar\pi_*^\cmm(t)} 
-\no{\psi^*_\cnn(t)\pi_*^\cmm(t)} )
 \Big\}.
\eens
In particular, the dangerous part
\be
\J_X = -e\int dt\ \sumnmp{} \mndmn X(q(t)) 
\no{\psi_\cnn(t)\pi^\cmm(t)}
\ee
now satisfies the algebra
\bes
[\J_X, \J_Y] &=& {k\/2\pi i} \int dt\ \dot X(q(t)) Y(q(t))
\nlb{JXYDir}
&=& {k\/2\pi i} \int dt\ \dot q^\mu(t)\dmu X(q(t)) Y(q(t)),
\eens
where the ``abelian charge'' is
\be
k = e^2 n {N+p\choose N},
\label{charge1}
\ee
and $n = 2^{[N/2]}$ is the number of spinor components.
The name ``abelian charge'' for the parameter in an abelian extension
was introduced in \cite{Lar98} in analogy with central charge for
central extensions. The extension (\ref{JXYDir}) is in fact central
if only the gauge generators are taken into account, but it does not
commute with arbitrary diffeomorphisms.

The regularized theory can now be quantized. We expand all jets 
($\psi_\cmm(t)$ etc.) in a Fourier series in $t$ and introduce a vacuum
$\ket{0}$ which is annihilated by all negative frequency modes. The
KT operator (\ref{QDir}) is already normal ordered and thus a
well-defined operator, so we can introduce dyna\-mics by passing to the
KT cohomology. Moreover, the theory so defined is a gauge theory on
the quantum level, because the gauge generators (\ref{JXDir}) act
in a well-defined manner on the history Hilbert space, and they
commute with the KT operator (\ref{QDir}):
\be
[\J_X, Q_{KT}] = 0.
\ee

However, making the gauge symmetry well-defined on the quantum level
comes with a price: the term on the RHS of (\ref{JXYDir}) is a anomaly.
This means that the quantum BRST charge (\ref{BRSTDir}) is no longer 
nilpotent.
To understand the physical implications of this, let us go back to the
classical level, and consider the action of the KT differential on
the Fourier modes of the fields.
\bes
\dlt \psi(k) &=& 0, \nl
\dlt \bar\psi(k) &=& 0, \nl
\dlt \psi^*(k) &=& (k_\mu\gm^\mu - M)\psi(k) 
- e\gm^\mu\int d^Nk' A_\mu(k-k')\psi(k'), \nl
\dlt \bar\psi^*(k) &=& \bar\psi(k)(k_\mu\gm^\mu + M) 
+ e\int d^Nk' A_\mu(k-k')\bar\psi(k')\gm^\mu, \nl
\dlt A_\mu(k) &=& 0, \\
\dlt \As^\mu(k) &=& k^\mu k_\nu A^\nu(k) - k^\nu k_\nu A^\mu(k) 
+ j^\mu(k), \nl
\dlt \zeta(k) &=& k_\mu \As^\mu(k) - 
ie\int d^Nk' (\bar\psi^*(k-k')\psi(k') + \bar\psi(k-k')\psi^*(k')).
\eens
If we ignore the non-linear terms proportional to $e$, the
physical phase space is spanned by fermions satisfying the Dirac
equation and massless photons. Unlike the free Maxwell case, however,
there are no ghosts $c$, and therefore there is one additional 
physical photon polarization. 
To see this in detail, let $k^2 = 0$, say $k = (k_0,0,0,k_0)$. 
Then $s A_\mu = 0$,
$s \As^0 = s \As^3 = k^0 k_\nu A^\nu$, $s\As^1 = s\As^2 = 0$ and
$s\zeta = k_\mu \As^\mu$. 
The kernel is thus spanned by $A_\mu$, $\As^1$, $\As^2$, and 
$k_\mu\As^\mu = \As^0 - \As^3$.
The image is spanned by $k^\mu A_\mu = A_0-A_3$ and $k_\mu\As^\mu$, which
factor out in cohomology. We are left with three physical polarizations
$A_1$, $A_2$, and $A_0+A_3$.

Three photon polarizations might seem counter-intuitive, and one might
wonder what happens with unitarity in this situation. This question is
beyond the scope of the present paper, since we have not even defined
the Hilbert space inner product. However, what must be
remembered is that we are dealing with the Hilbert space of the full,
interacting theory, and hence the photons are virtual. It is perfectly
legitimate for virtual photons to have unphysical polarizations, and this
is a necessary price for having a well-defined gauge symmetry on the
quantum level. The unphysical polarizations must vanish in regions of
spacetime where there is no charge. However, in that situation we are
dealing with a free Maxwell field, and the analysis in the previous
section applies.

Finally, we must address the question of removing the regularization,
i.e. to take the limit $p\to\infty$. This limit is problematic, because
the abelian charge (\ref{charge1}) diverges in all dimensions $N>0$.
This is perhaps not so surprising. We had to reject the field
formulation because the central extension in (\ref{charge0}) was
infinite. An infinite jet is essentially the same thing as the field
itself, and a physical divergence must therefore resurface in the jet
formalism.

However, the situation is not quite as bad as it seems. The antifields
also contribute to the abelian charge, but with opposite sign because
they have opposite Grassmann parity. Since the Dirac equation is first
order, the antifields are truncated at order $p-1$, so the total
abelian charge becomes
\be
k = e^2 n {N+p\choose N} - e^2 n {N+p-1\choose N} 
= e^2 n {N-1+p\choose N-1}.
\label{charge2}
\ee
This expression still diverges when $p\to\infty$ in all dimensions
$N>1$. Moreover, recall that we implicitly need further antifields
to cancel the $t$ dependence as in (\ref{dltsi}). After taking these
extra antifields, which are truncated at one order lower than the 
original fields and have opposite parity, the
abelian charge becomes
\be
k = e^2 n {N+p-1\choose N-1} - e^2 n {N+p-2\choose N-1} 
= e^2 n {N-2+p\choose N-2},
\label{charge3}
\ee
which has a finite limit when $N\leq2$.
For more complicated theories further cancellations
may be possible, pushing up the critical dimension.

It is important to realize that Maxwell's equations (\ref{EMax2})
and the continuity equation (\ref{cMax2}), or rather their jet
counterparts,
\bes
&&\sum_\nu ( F^{\mu\nu}_{\cmm+\nu}(t) - j^\mu_\cmm(t) ) = 0, \nle
&&\sum_\mu j^\mu_{\cmm+\mu}(t) \equiv 0,
\eens
are realized as operator equations in the KT cohomology.
This is ensured by the fact that the KT operator remains anomaly 
free after quantization, unlike the BRST operator.
Hence the extension (\ref{JXYDir}) of the algebra
of gauge transformations does not ruin charge conservation. 
This is possible because we do not identify the momentum $\pi(x)$
with the Dirac conjugate $\bar\psi(x)$, and hence the current 
(\ref{jmuDir}) is distinct from the gauge generators (\ref{JXpsi}).

In fact, without the anomaly it is impossible to write down a
well-defined operator for electic charge. By definition, the
electric charge operator $Q_{el}$ satisfies
\bes
&&[Q_{el}, \psi(x)] = -e\psi(x),  \qquad
[Q_{el}, \bar\psi(x)] = e\bar\psi(x), \nl
&&[Q_{el}, A_\mu(x)] = 0.
\ees
In jet space, this becomes
\bes
&&[Q_{el}, \psi_\cmm(t)] = -e\psi_\cmm(t),  \qquad
[Q_{el}, \bar\psi_\cmm(t)] = e\bar\psi_\cmm(t), \nl
&&[Q_{el}, A_{\mu\cmm}(t)] = 0.
\ees
Ignoring the action on the antifields, electric charge is 
thus measured by the operator
\be
Q_{el} = e \summ{p} \int dt\ \Big( 
\no{\bar\psi_\cmm(t)\bar\pi^\cmm(t)}
-\no{\psi_\cmm(t)\pi^\cmm(t)}
\Big).
\label{Qel}
\ee
This is recognized as the special gauge transformation
(\ref{JXDir}) $Q_{el} = \J_X$ with $X(x) \equiv 1$.

Since the electric charge is a gauge generator, it would be
impossible to have non-zero electric charge in the BRST cohomology.
In the KT cohomology, on the other hand, electric charge is a
conventional symmetry generator, because of the anomaly. 
It is easy to see that unitarity in fact requires the extension
in (\ref{JXYDir}). Namely, consider the restriction to the subalgebra
generated by $\J_X$ where $X(x) = X(x^0, 0, ..., 0)$ only depends
on the $x^0$ coordinate. This subalgebra is identified with the
affine algebra $\widehat{\uu(1)}$. It is well known that affine algebras
have non-trivial unitary representations only when the central 
charge is positive \cite{GO86}.
Since $Q_{el}\neq0$ means that the representation
is non-trivial, unitarity requires a non-zero anomaly.

\section{Yang-Mills theory}
\label{sec:Yang-Mills}

Non-abelian Yang-Mills theories are structurally quite similar to the
Maxwell theory studied in the previous sections, although the expressions
are more cumbersome. The main difference from our point of view is
that the gauge algebra becomes anomalous already for the pure theory,
because the quadratic Casimir no longer vanishes in the adjoint
representation. In other words, already pure Yang-Mills theory is an
interacting theory.

Let $\oj$ be a finite-dimensional Lie algebra with structure 
constants $f_{ab}{}^c$ and Killing metric $\dlt^{ab}$, which we
freely use to raise and lower $\oj$ indices. The gauge potential is
denoted by $A^a_\mu(x)$ and the exterior derivative reads
\be
\D_\mu = \dmu + iA^a_\mu(x)T_a,
\label{covder}
\ee
where the matrices $T_a$ belong to some finite-dimensional representation 
of $\oj$. The field strength is
\be
F_{\mu\nu}(x) = [\D_\mu, \D_\nu] = F^a_{\mu\nu}(x)T_a.
\ee
The action 
\be
S = \quart\int \dNx F^a_{\mu\nu}(x) F_a^{\mu\nu}(x)
\ee
leads to the Yang-Mills equations of motion,
\be
\D_\nu F_a^{\mu\nu}(x) = 0.
\ee
We introduce the KT and BRST differentials; the latter reads
\bes
s c^a(x) &=& -\half f_{bc}{}^a c^b(x)c^c(x), \nl
s A^a_\mu(x) &=& \D_\mu c^a(x), 
\nlb{sMax3}
s A_{*a}^\mu(x) &=& \D_\nu F_a^{\mu\nu}(x)
+ f_{ab}{}^c A_{*c}^\mu(x)c^b(x), \nl
s \zeta_a(x) &=& \D_\mu A_{*a}^\mu(x) 
+ f_{ab}{}^c \zc(x)c^b(x),
\eens
where
\be
\D_\mu c^a(x) = \d_\mu c^a - f_{bc}{}^a A^b_\mu(x)c^c(x),
\ee
in agreement with (\ref{covder})
After passage to jet space, the KT differential becomes
\bes
\dlt A^a_{\mu\cmm}(t) &=& 0,
\nle
\dlt A_{*a\cmm}^\mu(t) &=& (\D_\nu F_a^{\mu\nu})_\cmm(t), \nl
\dlt \zeta_{a\cmm}(t) &=& (\D_\mu A_{*a}^\mu)_\cmm(t).
\eens
The gauge generators
\bes
\J_X &=& \int dt\  \Big\{ 
\Big( \sumnmp{}\mndmn X^a(q(t))f_{ab}{}^c 
\no{ A^b_{\mu\cnn}(t)E^{\mu\cmm}_c(t)} \nl
&&\qquad+\summ{p}\sum_\mu \d_{\mm+\mu}X^a(q(t))E^{\mu\cmm}_a(t) \Big) 
\label{Jag} \\
&&-\sumnmp{-2} \mndmn  X^a(q(t))f_{ab}{}^c 
\no{ A^\mu_{*c\cnn}(t)E^{*b\cmm}_\mu(t) } \nl
&&- \sumnmp{-3} \mndmn X^a(q(t))f_{ab}{}^c 
\no{ \zeta_{c\cnn}(t)\chi^{b\cmm}(t) } \Big\}
\eens
satisfy the algebra
\bes
[\J_X, \J_Y] &=& \J_{[X,Y]} 
+ {k\/2\pi i} \int dt\ \dot X_a(q(t)) Y^a(q(t))
\nlb{JXYYM}
&=& \J_{[X,Y]} 
+ {k\/2\pi i} \dlt^{ab} \int dt\ \dot q^\mu(t)\dmu X_a(q(t)) Y_b(q(t)),
\eens
where the abelian charge has three contributions from the three sums
in (\ref{Jag}):
\be
k = f_{ac}{}^d f_{bd}{}^c \Big( {N+p\choose N} 
- {N+p-2\choose N} + {N+p-3\choose N} \Big).
\label{kg}
\ee
$k\neq 0$ because the quadratic Casimir $f_{ac}{}^d f_{bd}{}^c > 0$, so
there is an anomaly in the gauge algebra, which does not cancel the
contribution from the ghost $\cam(t)$. The BRST operator is not
nilpotent and only the KT cohomology can be implemented. In four 
dimensions, we have transverse gluons $A^a_1$ and $A^a_2$, but also
the polarization $A^a_0+A^a_3$ survives. This is not a
contradiction, because nonabelian Yang-Mills theory is an interacting
theory already without fermions, and virtual gluons need not be
transverse.
The passage to $p$-jet space is a regularization which can not be
removed. The leading divergences in the $p\to\infty$ limit of (\ref{kg})
do not cancel; that would require a term proportional to 
${N+p-1\choose N}$. 

We may ask under which condition the abelian charge $k$ has a finite 
$p\to\infty$ limit in four dimensions.
To that end, consider Yang-Mills theory coupled to a
fermionic spinor field $\psi$. Let $x_B$ and $x_F$ denote the
values of the quadratic Casimir for the bosonic and fermionic fields,
i.e. $\tr\ T_aT_b = x\dlt_{ab}$, and let $n_F$ denote the number of
fermionic species.
As in \cite{Lar02}, we find that the contributions from the fields
and the various antifields are
\bes
\barr{|c|c|c|l|}                                                       
\hline
\afn & \hbox{Jet} & \hbox{Order} &x \\
\hline
0 & \psi_\cmm(t) & p & n_Fx_F \\
1 & \psi^*_\cmm(t) & p-1 & -n_Fx_F \\
\hline
\hline
0 & A^a_{\mu\cmm}(t) & p & -x_B \\
1 & A^\mu_{*a\cmm}(t) & p-2 & x_B \\
2 & \zam(t) & p-3 & -x_B \\
\hline
\earr
\label{tab}
\ees
In order for the abelian charges to have a finite $p\to\infty$ limit,
the contributions at order $p-r$ in $N$ dimensions must be
\be
x_r = (-1)^r {N\choose r} X.
\ee
However, there are also antifields which cancel to $t$-dependence as
in (\ref{dltsi}), so the above equation is replaced by
\be
x_r = (-1)^r {N-1\choose r} X.
\ee
In particular, if we put $N=4$ and substitute the contributions from
(\ref{tab}), we find 
\bes
p: &\quad& n_Fx_F - x_B = X \nl
p-1: && -n_Fx_F = -3X, 
\nlb{4cond}
p-2: && x_B = 3X, \nl
p-3: && -x_B = -X.
\eens
These conditions clearly have no non-trivial solutions.

\section{Conclusion}
\label{sec:Conclusion}

In this paper we have extended the manifestly covariant canonical
quantization method, introduced in \cite{Lar04}, to theories with gauge
symmetries. The method is exact but implicit. We describe the
regularized Hilbert spaces exactly as cohomology spaces, but in order to
extract numbers we need a more explicit description. This presumably 
requires the introduction of perturbation theory and renormalization into
this formalism. 

The passage to $p$-jets is a regularization, and in the end we must
remove the regularization by taking the limit $p\to\infty$. It is clearly
necessary that all abelian charges remain finite in this limit. It was
hoped in \cite{Lar02,Lar03} that the leading infinities would cancel for
some particular field content describing our world. Under reasonable
assumptions, this requirement uniquely fixed spacetime dimension to $N=4$,
and vaguely suggested three generations of quarks and leptons, but the
details do not seem to work out. In particular, it is disappointing that
the conditions (\ref{4cond}) lack solutions. Perhaps this hints that
further renormalization is necessary.

To quantize regularized field theories might not seem overly
impressive. However, it should be emphasized that also the regularized
theories carry representations of the {\em full} constraint algebras, 
in the right number of dimensions, and that
this feature makes Taylor-expansion regularization unique. It is 
sometimes claimed that lattice gauge theory implements the gauge group
exactly, but this is not quite true.
The lattice gauge group is the group $Map(\Lambda, G)$ of maps from the
finite lattice $\Lambda$ into $G$, and this group is different from the 
continuum gauge group $Map(\RR^N, G)$. In particular, $Map(\Lambda, G)$
is finite-dimensional, so there are no gauge anomalies.

The main novelty is the appearence of new gauge anomalies.
This phenomenon is genuinely new, because neither a Yang-Mills anomaly
proportional to the quadratic Casimir, nor a pure diffeomorphism anomaly
in four dimensions, can arise in conventional quantum field theory
\cite{Bon86}. The anomaly vanishes in the particular case of the free
Maxwell field, because the second Casimir is zero in the adjoint
representation of $\uu(1)$, but it is generically present in interacting
theories. 

Describing physics as cohomology in the history phase space is very
convenient, because it maintains manifest covariance at all times.
Indeed, the history phase space is the natural habitat for the algebras
of gauge transformations and diffeomorphisms. However, it is presumably
possible to repeat the analysis, including the anomalies, using a
privileged foliation of spacetime. In constrast, to introduce the
observer's trajectory is essential, because otherwise it is impossible
to even write down the new, observer-dependent anomalies.

We also argued that these new anomalies do not imply any inconsistency,
unlike conventional gauge anomalies caused by chiral fermions coupled to
gauge fields. On the contrary, consistency (unitarity) is guaranteed by
unitary representations, and all non-trivial unitary irreps of gauge and
diffeomorphism algebras are projective. In particular, unitary time
evolution is guaranteed if the Hamiltonian is an element in a unitarily
represented algebra, anomalous or not. Moreover, a non-zero electric
charge (\ref{Qel}), which certainly is a necessary physical requirement,
can only be a well-defined operator in the presence of the anomaly. It
would be inconsistent, however, to try to treat an anomalous gauge
symmetry as a gauge symmetry. Rather, the anomaly makes classical gauge
degrees of freedom physical on the quantum level, turning the classical
gauge symmetry into a conventional global quantum symmetry. The 
physical consequence is typically that
virtual photons and gluons may have unphysical polarizations.

Despite the phenomenological success of quantum field theory, these new
anomalies can not be ignored in a final theory, simply because they
exist mathematically. Anomalies matter!

\end{document}